\pdfoutput=1

\documentclass[11pt,twoside,a4paper,cmspaper,final,collab]{cms-tdr}
\def\svnVersion{166200}\def\cmsCernNo{2011-222}\def\cmsCernDate{\today}\def\cmsMessage{Submitted to the European Physical Journal C}

\begin{document}\cmsNoteHeader{HIN-11-006}

\hyphenation{had-ron-i-za-tion}
\hyphenation{cal-or-i-me-ter}
\hyphenation{de-vices}

\RCS$Revision: 117473 $
\RCS$HeadURL: svn+ssh://svn.cern.ch/reps/tdr2/papers/HIN-11-006/trunk/HIN-11-006.tex $
\RCS$Id: HIN-11-006.tex 117473 2012-04-21 01:43:25Z alverson $
\newcommand {\roots}    {\ensuremath{\sqrt{s}}}
\newcommand {\rootsNN}  {\ensuremath{\sqrt{s_{_{NN}}}}}
\newcommand {\pttrg}       {\ensuremath{p_\mathrm{T}^{\mathrm{trig}}}}
\newcommand {\ptass}       {\ensuremath{p_\mathrm{T}^{\mathrm{assoc}}}}
\newcommand {\ptlow}       {\ensuremath{p_\mathrm{T}^{\mathrm{low}}}}
\newcommand {\npart}    {\ensuremath{\mathrm{N}_\mathrm{part}}}

\def\d{\mathrm{d}}
\providecommand{\titlerunning}{\relax} 

\cmsNoteHeader{HIN-11-006} 
\title{Centrality dependence of dihadron correlations and azimuthal anisotropy harmonics in PbPb collisions at \texorpdfstring{$\rootsNN = 2.76\TeV$}{sqrt(s[NN]) = 2.76 TeV}}

\date{\today}

\abstract{
Measurements from the CMS experiment at the LHC
of dihadron correlations for charged particles produced in PbPb collisions
at a nucleon-nucleon centre-of-mass energy of 2.76\TeV are presented.
The results are reported as a function of the particle transverse momenta (\pt)
and collision centrality over a broad range in relative pseudorapidity ($\Delta\eta$)
and the full range of relative azimuthal angle ($\Delta\phi$).
The observed two-dimensional
correlation structure in $\Delta \eta$ and $\Delta \phi$ is characterised
by a narrow peak at $(\Delta\eta, \Delta\phi) \approx (0, 0)$ from jet-like
correlations and a long-range structure that persists up to at least $|\Delta\eta| = 4$.
An enhancement of the magnitude of the short-range jet peak is observed
with increasing centrality, especially for particles of \pt\ around 1--2\GeVc.
The long-range azimuthal dihadron correlations are extensively studied
using a Fourier decomposition analysis.
The extracted Fourier coefficients are found to factorise into a product
of single-particle azimuthal anisotropies up to \pt\ $\approx$ 3--3.5\GeVc for at least one particle
from each pair, except for the second-order harmonics in the most central PbPb events.
Various orders of the single-particle azimuthal anisotropy harmonics
are extracted for associated particle \pt\ of 1--3\GeVc, as a
function of the trigger particle \pt\ up to 20\GeVc and over the full centrality
range.
}

\hypersetup{%
pdfauthor={CMS Collaboration},%
pdftitle={Centrality dependence of dihadron correlations and azimuthal anisotropy harmonics in PbPb collisions at sqrt(s[NN]) = 2.76 TeV},%
pdfsubject={CMS},%
pdfkeywords={CMS, correlations, dihadron, ridge, factorization, fourier}}
\titlerunning{Dihadron correlations and azimuthal anisotropies in PbPb collisions} 
\maketitle 

\section{Introduction}
\label{sec:intro}

Measurements of dihadron correlations are a well established technique for
studying the properties of particle production in the high density medium
created in heavy ion collisions.
Early results from PbPb collisions at the Large Hadron Collider
(LHC)~\cite{ref:HIN-11-001-PAS,Collaboration:2011by}
extended these studies into a regime of much higher beam energies as compared
to those from the Relativistic Heavy Ion Collider (RHIC)
~\cite{Adler:2002tq,star:2009qa,star:2010ridge,Adams:2005ph,Adare:2006nr,phenix:2008cqb, Alver:2009id,Alver:2008gk}.
These results complement other LHC measurements of medium properties, including
a large deficit of charged particles at high-\pt~\cite{Aamodt:2010jd} and
the observations of an enhanced fraction of dijets with very asymmetric
energies~\cite{Aad:2010bu,HIN-10-004}.

The Compact Muon Solenoid (CMS) experiment at the LHC has studied dihadron correlations
over a broad range of relative azimuthal angles ($|\Delta\phi|$) and pseudorapidity
($|\Delta\eta|$, where $\eta = -\ln[\tan(\theta/2)]$ and $\theta$ is the polar angle
relative to the counterclockwise beam axis) in the most central PbPb collisions
at a nucleon-nucleon centre-of-mass energy (\rootsNN) of 2.76\TeV~\cite{ref:HIN-11-001-PAS}.
Concentrating on large $|\Delta\eta|$, previous measurements at RHIC established some of
the properties of the so-called ``ridge''~\cite{Adams:2005ph,star:2009qa,Alver:2009id},
an enhancement of pairs with $|\Delta\phi| \approx 0$. While a variety of
theoretical models have been proposed to interpret the ridge
phenomena as a consequence of jet-medium interactions~\cite{Armesto:2004pt,
Majumder:2006wi,Chiu:2005ad,Wong:2008yh,Romatschke:2006bb,Shuryak:2007fu},
recent theoretical developments indicate that, because of event-by-event
fluctuations in the initial shape of the interacting region, sizeable
higher-order hydrodynamic flow terms could also be induced, e.g.,
triangular flow~\cite{Voloshin:2004th,Mishra:2007tw,Takahashi:2009na,Alver:2010gr,Alver:2010dn,Schenke:2010rr,
Petersen:2010cw,Xu:2010du,Teaney:2010vd}. The triangular flow effect
will contribute to the dihadron correlations in the form of a $\cos(3\Delta\phi)$
component, which also gives a maximum near-side correlation at $\Delta\phi \approx 0$,
similarly to the elliptic flow contribution. It has been proposed that by
taking into account various higher-order terms, the ridge structure
could be described entirely by hydrodynamic flow effects~\cite{Alver:2010gr}.
To investigate this possibility, a Fourier decomposition of the CMS data at
large $|\Delta\eta|$ was performed, finding a strong
dependence on \pt~\cite{ref:HIN-11-001-PAS}.
Similar results, although with a smaller $|\Delta\eta|$ gap, have been reported
by ALICE~\cite{Collaboration:2011by,ALICE:2011ab}.
The observations by CMS of a ridge-like structure in very high multiplicity proton-proton
(pp) collisions at a centre-of-mass energy of 7\TeV~\cite{Khachatryan:2010gv},
where no medium effect is expected, may also challenge the
interpretations of these long-range correlations.

This paper, expanding on previous CMS results~\cite{ref:HIN-11-001-PAS}, presents dihadron
correlation measurements from PbPb collisions at \rootsNN\ = 2.76\TeV acquired in 2010,
for all collision centralities and
over a broader range of hadron \pt. As in Refs.~\cite{Khachatryan:2010gv,ref:HIN-11-001-PAS}, the
yield of particles (binned in \pt) associated with a trigger particle
(also binned in \pt) is extracted as a function of their relative pseudorapidity
and azimuthal angle.
Such a study of hadron pairs in either the same or different
\pt\ ranges can reveal important information about the production
of particles and their propagation through the medium.
A Fourier decomposition technique is used to quantify the
long-range azimuthal correlations. The potential connection
between the extracted Fourier coefficients from the correlation
data and the azimuthal anisotropy harmonics for single particles
is investigated. This measurement provides a comprehensive
examination of the centrality and transverse momentum
($1<\pt<20$\GeVc) dependencies of the short-range
($|\Delta\eta| < 1$) and long-range ($2 < |\Delta\eta| < 4$)
dihadron correlations in PbPb collisions at LHC energies,
as well as the relationship between these two-particle
correlations and single-particle angular distributions.
These results provide extensive input to the interpretation
of these observables in terms of broad theoretical concepts
such as hydrodynamic flow and quantitative models of
particle production and propagation in the high-density medium.

The detector, the event selection and the extraction of the
correlation functions are described in Sections~\ref{sec:detector}
and~\ref{sec:analysis}, while the extracted results are described
in Sections~\ref{sec:correlation} and~\ref{sec:PbPbFourierDecompositionAnalysis}.

\section{CMS Detector}
\label{sec:detector}
The ability of the CMS detector (all components of which are described
in Ref.~\cite{JINST}) to extract the properties of charged
particles over a large solid angle is particularly important in the
study of dihadron correlations. This study is based primarily on data from
the inner tracker contained within the 3.8\unit{T} axial magnetic field of the large
superconducting solenoid. The tracker consists of silicon pixel and strip
detectors. The
former includes 1\,440  modules arranged in 3 layers, while
the latter consists of 15\,148 modules arranged in 10 (11) layers in the
barrel (endcap) region. The trajectories of charged particles can be
reconstructed for $\pt > 100$\MeVc and within $|\eta| < 2.5$.

The field volume of the solenoid also contains crystal electromagnetic
and brass/scintillator hadron calorimeters. Although not included in
the present results, muons are detected using gas-ionisation counters
embedded in the steel return yoke. In addition to these components in
and around the barrel and endcap of the solenoid, CMS also has extensive
forward calorimetry. In the right-handed coordinate system used by CMS,
the  $x$-, $y$-, and $z$-axes are aligned with the radius of the
LHC ring, the vertical direction, and the counterclockwise beam
direction, respectively, with the origin located at the centre of
the nominal interaction region.

For PbPb collisions, the primary minimum-bias trigger uses signals from
either the  beam scintillator counters (BSC, $3.23 < |\eta| < 4.65$) or the
steel/quartz-fibre Cherenkov forward hadron calorimeters
(HF, $2.9 < |\eta| < 5.2$). Coincident signals from
detectors located at both ends of the detector (i.e., a pair of BSC or a pair
of HF modules) are required. Events due to noise, cosmic-ray muons, double-firing
triggers, and beam backgrounds are suppressed by further requiring the presence
of colliding beam bunches. The fraction of  inelastic hadronic PbPb collisions
accepted by this primary trigger is $(97 \pm 3)$\%~\cite{HIN-10-004}.

\section{Data and Analysis}
\label{sec:analysis}
The procedure used in the present analysis follows that described in
the previous CMS correlation paper~\cite{ref:HIN-11-001-PAS}. Offline event
selection requires a reconstructed vertex with at least two tracks (i.e., at
least one pair of charged particles). This vertex must be within 15~cm along
the beam axis relative to the centre of the nominal collision region and
within 0.02~cm in the transverse plane relative to the average position of
all vertices in a given data sample. In addition, various background events
(for example  beam-gas and beam-halo collisions, cosmic muons, and
large-impact-parameter electromagnetic collisions) are suppressed by
requiring at least three signals in the HF calorimeters at both positive and negative $\eta$,
with at least 3\GeV of energy in each signal.

The analysis is based on a data sample of PbPb collisions corresponding
to an integrated luminosity of approximately 3.9~$\mu$b$^{-1}$~\cite{EWK-10-004,EWK-11-001},
which contains 30~million minimum-bias collisions after all event
selections are applied. The pp data at \roots\ = 2.76\TeV, the
reference for comparison to the PbPb data, were collected during a short
low-energy LHC run at the end of March 2011.
Minimum-bias-triggered pp events corresponding to an integrated luminosity of 520~$\mu$b$^{-1}$
are selected for this analysis.

The energy released in the collisions is related to the centrality
of the heavy ion interactions, i.e., the geometrical overlap
of the incoming nuclei. The event centrality is defined
as the fraction of the total cross section, starting at 0\% with
the most central collisions (i.e.,\ smallest impact parameter). This fraction
is determined from the distribution of total energy measured
in both HF calorimeters.

The event centrality can be correlated with the total number
of nucleons in the two Pb nuclei that experienced at least
one inelastic collision, \npart. The average values of \npart\
for the various centrality bins used in this analysis are given in Table~\ref{table:npart}.
The \npart\ values are obtained using a Glauber Monte Carlo (MC)
simulation~\cite{glauber,Alver:Glauber} with the same parameters
as in Ref.~\cite{HIN-10-004}. These calculations are translated
into reconstructed centrality bins using correlations between
\npart\ and the measured total energy in the HF calorimeters,
obtained from fully simulated MC events. The systematic uncertainties
on the \npart\ values in Table~\ref{table:npart} are derived from propagation of the
uncertainties in the parameters of the Glauber model. More
details on the determination of centrality and \npart\ can be
found in Refs.~\cite{Chatrchyan:2011pb,HIN-10-004,D'Enterria:2007xr}.

\begin{table*}[htb]
\centering
\topcaption{\label{table:npart} Average \npart\ values for each PbPb centrality range used in this paper.
The values are obtained using a Glauber MC simulation with the same parameters as in
Ref.~\cite{HIN-10-004}.}
\begin{tabular}{c|cccccccccccccccccc}
\hline
\hline
Centrality & 0--5\% & 5--10\% & 10--15\% & 15--20\% & 20--25\% & 25--30\% \\
$\langle$\npart$\rangle$ & $381\pm2$ & $329\pm3$ & $283\pm3$ & $240\pm3$ & $203\pm3$ & $171\pm3$ \\
\hline
Centrality & 30--35\% & 35--40\% & 40--50\% & 50--60\% & 60--70\% & 70--80\% \\
$\langle$\npart$\rangle$ & $142\pm3$ & $117\pm3$ & $86.2\pm2.8$ & $53.5\pm2.5$ & $30.5\pm1.8$ & $15.7\pm1.1$ \\
\hline
\hline
\end{tabular}
\end{table*}

The reconstruction of charged particles in PbPb collisions is based
on signals in the silicon pixel and strip detectors, similarly
to the reconstruction for pp collisions~\cite{TRK-10-001}.
However, a number of settings are adjusted to cope
with the challenges presented by the much higher signal density
in central PbPb collisions.
A set of tight quality selections are imposed
on the collection of fully reconstructed tracks
to minimise the contamination from misidentified tracks.
These include requirements of at least 13 signals on the track, a relative momentum
uncertainty of less than 5\%, a normalised $\chi^2$ of less than
0.15 times the number of signals, and transverse and longitudinal
impact parameters of less than three times the sum in quadrature
of the uncertainties on the impact parameter and the primary
vertex position. Studies with simulated MC events show that
the combined geometrical acceptance and reconstruction efficiency
for the primary-track reconstruction reaches about 60\% for the 0--5\%
most central PbPb collisions at $\pt > 2\GeVc$ over the full CMS
tracker acceptance ($|\eta| < 2.4$) and 66\% for $|\eta| < 1.0$.
The fraction of misidentified tracks is about 1--2\%
for $|\eta| < 1.0$, but increases to 10\% at $|\eta| \approx 2.4$
for the 5\% most central PbPb collisions. For the peripheral PbPb events (70--80\%),
the overall tracking efficiency improves by up to 5\%, with a much lower
fraction of misidentified tracks.

The analysis of dihadron angular correlations in this paper follows
exactly the procedure established in Ref.~\cite{ref:HIN-11-001-PAS}.
Any charged particle associated with the primary vertex and
in the range $|\eta| < 2.4$ can be used as a trigger particle.
A variety of bins of trigger transverse momentum, denoted by \pttrg, are considered.
There can be more than one such trigger particle in a single event
and their total multiplicity in a particular data sample is denoted by $N_\text{trig}$.
Within each event, every trigger particle is then paired
with all of the remaining particles (again within  $|\eta| < 2.4$).
As for the trigger particles, these associated particles
are binned in transverse momentum (\ptass).
The differential yield of associated particles per trigger particle is given by
\begin{linenomath}
\begin{equation}
\label{2pcorr_incl}
\frac{1}{N_\text{trig}}\frac{\d^{2}N^\text{pair}}{\d\Delta\eta\, \d\Delta\phi}
= B(0,0)\times\frac{S(\Delta\eta,\Delta\phi)}{B(\Delta\eta,\Delta\phi)},
\end{equation}
\end{linenomath}
where $N^\text{pair}$ is the total number
of correlated hadron pairs. The functions $S(\Delta\eta,\Delta\phi)$ and
$B(\Delta\eta,\Delta\phi)$ are called the signal and background distributions,
respectively. The value of the latter at $\Delta\eta=0$ and $\Delta\phi=0$ ($B(0,0)$)
is a normalisation factor.

The signal distribution is the per-trigger-particle yield of pairs found in the same event,
\begin{linenomath}
\begin{equation}
\label{eq:signal}
S(\Delta\eta,\Delta\phi) = \frac{1}{N_\text{trig}}\frac{\d^{2}N^\text{same}}{\d\Delta\eta\, \d\Delta\phi},
\end{equation}
\end{linenomath}
where $N^\text{same}$ is the number of such pairs within a ($\Delta\eta$,$\Delta\phi$) bin.
The background distribution is found using a mixed-event technique, wherein trigger
particles from one event are combined (mixed) with all of the associated particles from
a different event. In the analysis, associated particles from 10 randomly chosen
events are used. The result is given by
\begin{linenomath}
\begin{equation}
\label{eq:background}
B(\Delta\eta,\Delta\phi) = \frac{1}{N_\text{trig}}\frac{\d^{2}N^\text{mix}}{\d\Delta\eta\, \d\Delta\phi},
\end{equation}
\end{linenomath}
where $N^\text{mix}$ denotes the number of
mixed-event pairs. This background distribution represents the expected correlation
if the only effects present were random combinatorics and pair-acceptance.

The value of $B(\Delta\eta,\Delta\phi)$ at $\Delta\eta=0$ and $\Delta\phi=0$
(with a bin width of 0.3 in $\Delta\eta$ and $\pi/16$ in $\Delta\phi$) is used
to find the normalisation factor $B(0,0)$.
In this case, the two particles have the maximum possible geometric pair acceptance
since they are travelling in essentially the same direction.
The effect of two tracks merging into a single reconstructed track is negligible.
The extent to which the background distribution at larger angular separation
is smaller than this value (more specifically the ratio \ifthenelse{\boolean{cms@external}}{\linebreak[4]}{}$B(0,0)/B(\Delta\eta,\Delta\phi)$)
can be used to determine the pair acceptance
correction factor. Multiplying the signal distribution by this ratio
gives the acceptance-corrected per-trigger-particle associated
yield. Since the distributions should, in principle, be symmetric, the statistical precision
is maximised by filling only one quadrant using the absolute values of $\Delta\eta$
and $\Delta\phi$. For illustration purposes only (for example, see
Fig.~\ref{fig:Corr2D_T1Min30T1Max35_A1Min10A1Max15_FlowOrder0}),
the other three quadrants are filled by reflection, giving distributions that
are symmetric about $(\Delta\eta,\Delta\phi) = (0,0)$ by construction.
The pair acceptance decreases rapidly with $\Delta\eta$ and so, to avoid large
fluctuations due to statistical limitations, the distributions are
truncated at $|\Delta\eta| = 4$.
The analysis is performed in twelve centrality classes of PbPb
collisions ranging from the most central 0--5\% to the most peripheral
70--80\%. Within each centrality range, the yield described in
Eq.~\ref{2pcorr_incl} is calculated in 0.5~cm wide bins of the vertex
position ($z_\mathrm{vtx}$) along the beam direction and then averaged over the range
$|z_\mathrm{vtx}| < 15\cm$.

When filling the signal and background distributions, each pair is weighted by the
product of correction factors for the two particles. These factors
are the inverse of an efficiency that is a function of each particle's
pseudorapidity and transverse momentum,
\begin{linenomath}
\begin{equation}
\varepsilon_\text{trk}(\eta,\pt) = \frac{A(\eta,\pt) E(\eta,\pt)}{1-F(\eta,\pt)},
\end{equation}
\end{linenomath}
where $A(\eta,\pt)$ is the geometrical acceptance, $E(\eta,\pt)$
is the reconstruction efficiency, and $F(\eta,\pt)$ is the fraction of
misidentified tracks. The effect of this weighting factor only changes
the overall scale but not the shape of the associated yield distribution,
which is determined by the signal-to-background ratio.

As described in Ref.~\cite{ref:HIN-11-001-PAS}, the track-weighting procedure is tested
using MC events generated with {\sc HYDJET}~\cite{Lokhtin:2005px} (version 1.6) propagated
through a full detector simulation. The tracking efficiencies themselves are checked
using simulated tracks embedded into actual data
events. Systematic uncertainties due to variations of the track reconstruction efficiency
as a function of vertex location and also the procedure used to
generate the background events are evaluated.
The individual contributions are added in quadrature to find the final
systematic uncertainties of 7.3--7.6\%.

\section{Correlation Functions and Near-Side Yields}
\label{sec:correlation}

\begin{figure*}[thbp]
  \begin{center}
    \includegraphics[width=0.9\textwidth]{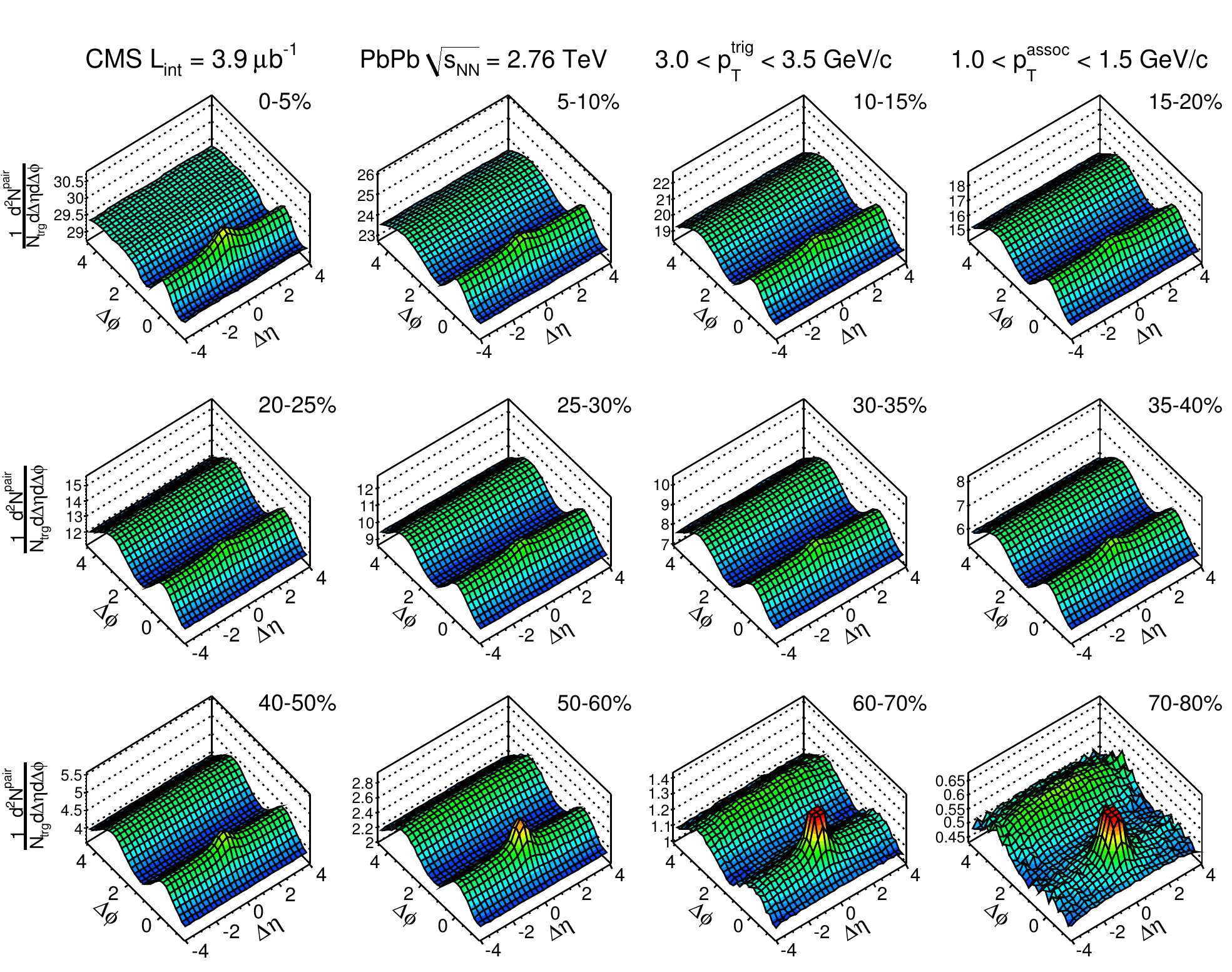}
    \caption{
         Two-dimensional (2D) per-trigger-particle associated yield of charged hadrons
         as a function of $|\Delta\eta|$ and $|\Delta\phi|$ for
         $3<\pttrg<3.5$\GeVc and $1<\ptass<1.5$\GeVc, for
         twelve centrality ranges of PbPb collisions at \rootsNN\ = 2.76\TeV.
         The near-side peak is truncated in the two most peripheral distributions to better display the
         surrounding structure.}
    \label{fig:Corr2D_T1Min30T1Max35_A1Min10A1Max15_FlowOrder0}
  \end{center}
\end{figure*}

The two-dimensional (2D) per-trigger-particle associated yield
distribution of charged hadrons as a function of $|\Delta\eta|$
and $|\Delta\phi|$ is measured for each \pttrg\ and \ptass\
interval, and in different centrality classes of PbPb collisions.
An example for trigger particles
with $3<\pttrg<3.5\GeVc$ and associated particles with $1<\ptass<1.5\GeVc$
is shown in Fig.~\ref{fig:Corr2D_T1Min30T1Max35_A1Min10A1Max15_FlowOrder0},
for centralities ranging from the 0--5\% most central collisions,
to the most peripheral (70--80\%)
events. The 2D correlations
are rich in structure, and evolve with centrality. The \pttrg\
and \ptass\ ranges shown in this figure were chosen as an example
because they demonstrate a good balance of the following features.
For the most central PbPb collisions, a clear and significant
ridge-like structure mostly flat in $\Delta\eta$, and extending
to the limit of $|\Delta\eta| = 4$, is observed
at $\Delta\phi \approx 0$.
At mid-peripheral events, a pronounced $\cos(2\Delta\phi)$
component emerges, originating predominantly from elliptic flow~\cite{Alver:2008gk}.
Lastly, in the most peripheral collisions, the near-side ridge structure
has largely diminished, while the away-side back-to-back jet correlations can be
clearly seen at $\Delta\phi \approx \pi$, but spread out in $\Delta\eta$.

\begin{figure*}[thbp]
  \begin{center}
    \includegraphics[width=1.\textwidth]{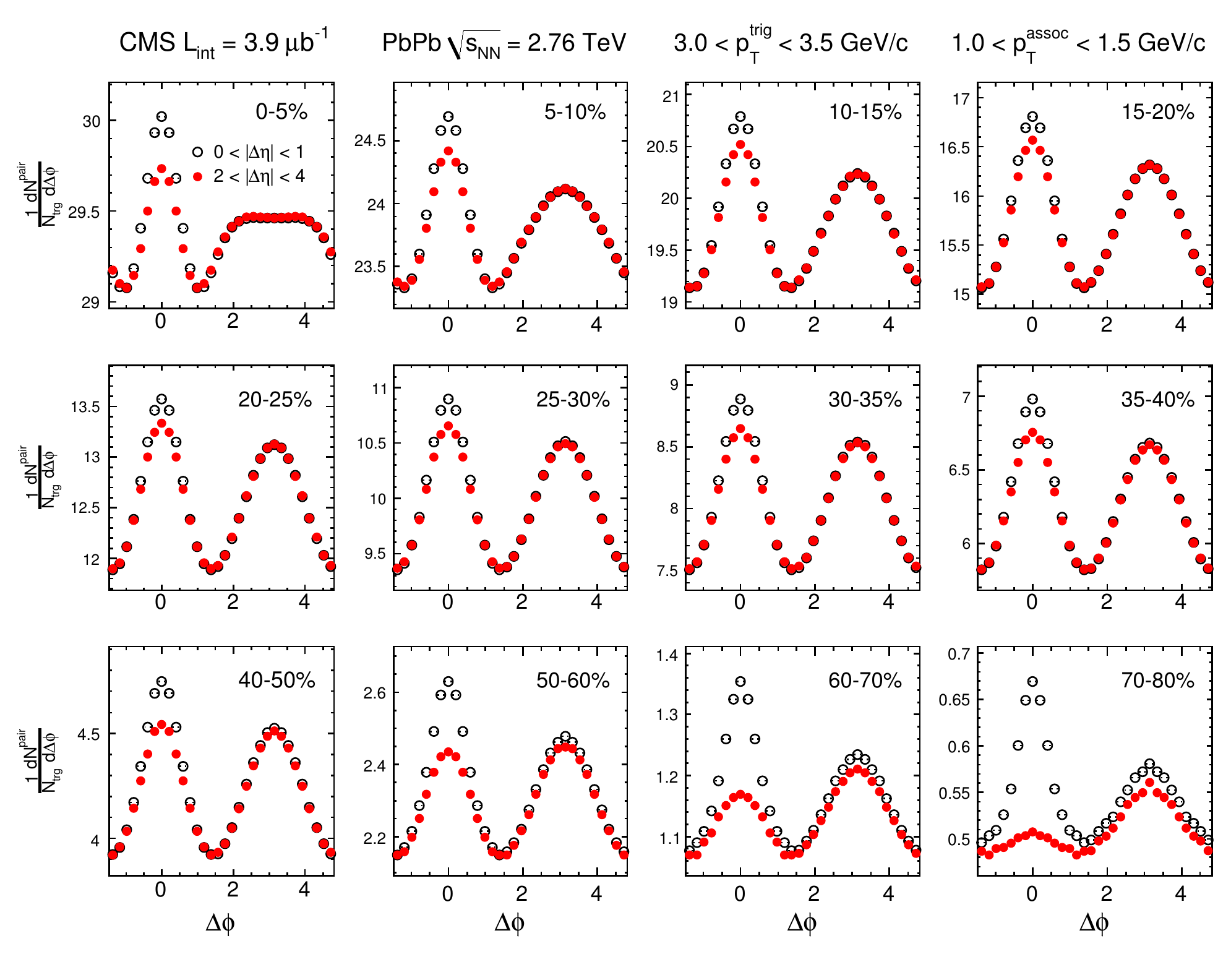}
    \caption{ Short-range ($0<|\Delta\eta|<1$, open circles) and long-range
         ($2<|\Delta\eta|<4$, red closed circles) per-trigger-particle associated
         yields of charged hadrons as a function of $|\Delta\phi|$ for
         $3<\pttrg<3.5$\GeVc and $1<\ptass<1.5$\GeVc, for twelve centrality
         ranges of PbPb collisions at \rootsNN\ = 2.76\TeV.
         The statistical error bars are smaller than the marker size.
         The systematic uncertainties of 7.6\% for all
         data points in the short-range region and 7.3\% for all data points
         in the long-range region are not shown in the plots.
          }
    \label{fig:Correlations1D_Jet_And_Ridge_Region}
  \end{center}
\end{figure*}

As was done in Ref.~\cite{ref:HIN-11-001-PAS}, to quantitatively
examine the features of short-range and long-range
azimuthal correlations, one-dimensional (1D) $\Delta\phi$
correlation functions are calculated by averaging the 2D distributions
over a limited region in $\Delta\eta$ from $\Delta\eta_\text{min}$ to $\Delta\eta_\text{max}$:
\ifthenelse{\boolean{cms@external}}{
\begin{multline}
\label{2pcorr_incl_1D}
\frac{1}{N_\text{trig}}\frac{\d N^\text{pair}}{\d\Delta\phi}
= \\ \frac{1}{\Delta\eta_\text{max}-\Delta\eta_\text{min}}
\int_{\Delta\eta_\text{min}}^{\Delta\eta_\text{max}}
\frac{1}{N_\text{trig}}\frac{\d^{2}N^\text{pair}}{\d\Delta\eta\, \d\Delta\phi}\d\Delta\eta.
\end{multline}
}{
\begin{linenomath}
\begin{equation}
\label{2pcorr_incl_1D}
\frac{1}{N_\text{trig}}\frac{\d N^\text{pair}}{\d\Delta\phi}
= \frac{1}{\Delta\eta_\text{max}-\Delta\eta_\text{min}}
\int_{\Delta\eta_\text{min}}^{\Delta\eta_\text{max}}
\frac{1}{N_\text{trig}}\frac{\d^{2}N^\text{pair}}{\d\Delta\eta\, \d\Delta\phi}\d\Delta\eta.
\end{equation}
\end{linenomath}
}
The results of extracting the 1D $\Delta\phi$ correlations in the
short-range ($0 < |\Delta\eta| < 1$) and long-range ($2 < |\Delta\eta| < 4$)
regions are shown in Fig.~\ref{fig:Correlations1D_Jet_And_Ridge_Region}.
The associated yield distribution per trigger particle
is extracted for the same \pttrg\ and \ptass\ ranges as
in Fig.~\ref{fig:Corr2D_T1Min30T1Max35_A1Min10A1Max15_FlowOrder0}.

\begin{figure*}[thbp]
  \begin{center}
    \includegraphics[width=1.0\textwidth]{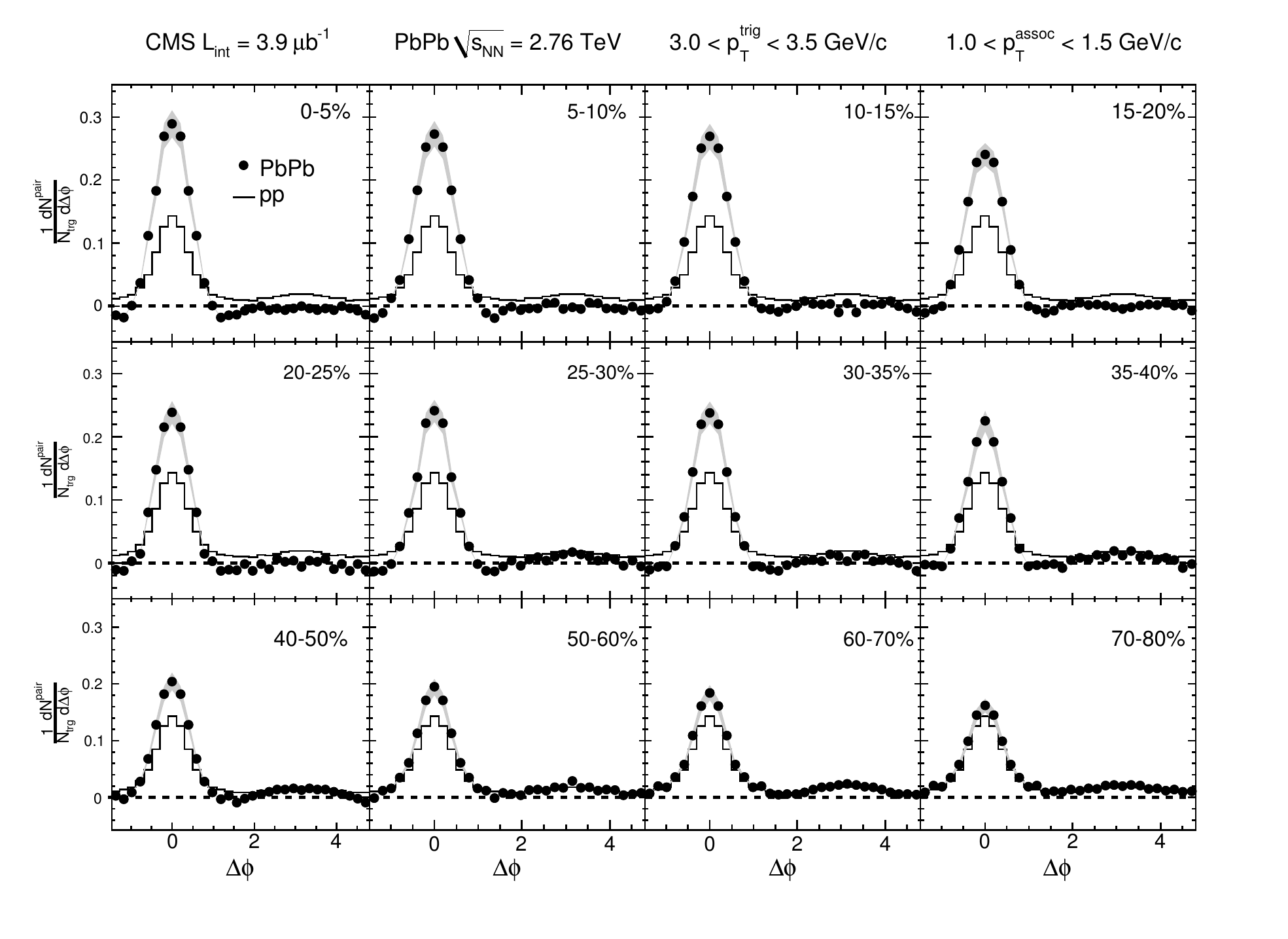}
    \caption{
         The difference between short-range ($0<|\Delta\eta|<1$) and long-range
         ($2<|\Delta\eta|<4$) per-trigger-particle associated
         yields of charged hadrons as a function of $|\Delta\phi|$ for
         $3<\pttrg<3.5$\GeVc and $1<\ptass<1.5$\GeVc, for twelve centrality
         ranges of PbPb collisions at \rootsNN\ = 2.76\TeV.
         The statistical error bars are smaller than the marker size.
         The grey bands denote the systematic uncertainties.
         The pp result is superimposed in all panels, for reference purposes.
          }
    \label{fig:Correlation1D_12Cent_JetMinusRidge}
  \end{center}
\end{figure*}

In order to study the short-range $\Delta\phi$ correlations in the absence of
the flat background in $\Delta\eta$, the 1D $\Delta\phi$ distribution
in the long-range region is subtracted from that in the short-range
region. The resulting difference of the distributions is shown
in Fig.~\ref{fig:Correlation1D_12Cent_JetMinusRidge}. The near-side
peak ($\Delta\phi \approx 0$) represents mainly the correlations
from jet fragmentation, whereas the away-side region
($\Delta\phi \approx \pi$) is mostly flat and close to zero due to
the weak $\Delta\eta$ dependence of the away-side jet peak.
A comparison to the pp data at \roots\ = 2.76\TeV
is also presented, showing a similar structure to that in the very
peripheral 70--80\% PbPb data. However, the magnitude of the near-side
peak is significantly enhanced in the most central PbPb collisions as compared
to pp. Most of the systematic uncertainties 
manifest themselves as an overall change in the scale of the correlation functions,
with little dependence on $\Delta\phi$ and $\Delta\eta$. Therefore, they
largely cancel when the difference between the short-range and long-range
regions is taken.

The strengths of the near-side peak and away-side region
in the $|\Delta\phi|$ distributions from Fig.~\ref{fig:Correlation1D_12Cent_JetMinusRidge}
can be quantified by integrating over the two $|\Delta\phi|$
ranges separated by the minimum position of the distribution.
This position is chosen as $|\Delta\phi| = 1.18$, the
average of the minima between the near side and away side
over all centralities. This choice of integration range introduces
an additional systematic uncertainty in the integrated associated yields.
The effect of choosing different minima for integration ranges changes the overall
yield by an absolute amount of at most 0.007.
Similar shifts are calculated for each data point, then
added as an absolute value in quadrature to the other systematic
uncertainties.

\begin{figure*}[thbp]
  \begin{center}
    \includegraphics[width=\textwidth]{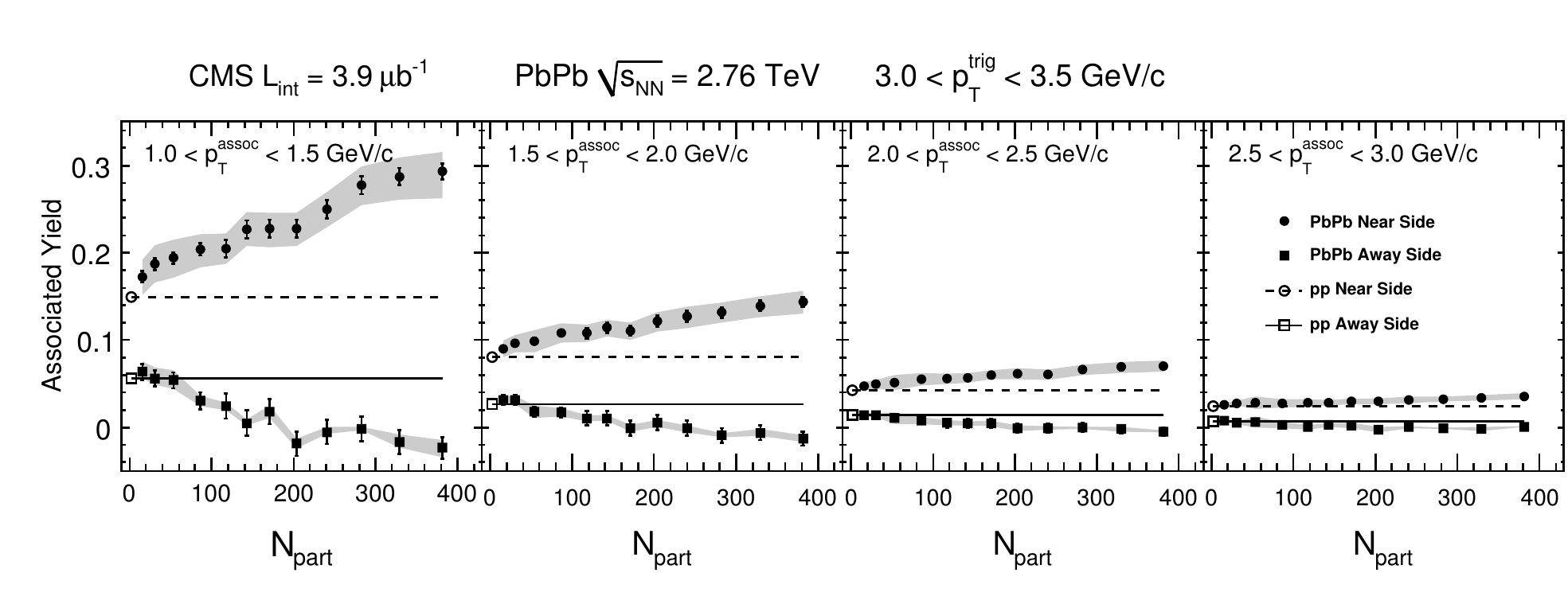}
    \caption{ The integrated associated yields of the near-side peak
    ($|\Delta\phi|<1.18$) and away-side region ($|\Delta\phi|>1.18$),
    requiring $3 < \pttrg < 3.5\GeVc$ for four different intervals of
    \ptass, as a function of \npart\ in PbPb collisions at
    \rootsNN\ = 2.76\TeV. The statistical error bars are smaller
    than the marker size. The grey bands denote the systematic uncertainties.
   The lines represent the pp results (\npart\ = 2) superimposed over the full range of \npart\ values.
    }
    \label{fig:Correlation1D_4Panel_Yields_TrigPtBin_4}
  \end{center}
\end{figure*}

Figure~\ref{fig:Correlation1D_4Panel_Yields_TrigPtBin_4} shows
the integrated associated yields of the near-side peak and
away-side regions as a function of \npart\ in PbPb collisions at
\rootsNN\ = 2.76\TeV, requiring $3 < \pttrg < 3.5\GeVc$, for four
different intervals of \ptass\ (1--1.5, 1.5--2, 2--2.5, and 2.5--3\GeVc). The
grey bands represent the systematic uncertainties.
For easier visual comparison between the most central PbPb
results and the values one would expect from a trivial
extrapolation of the pp results, the latter are also
represented using horizontal lines covering the full \npart\ range.
The yield of the near-side peak increases by a factor of 1.7
in going from the very peripheral 70--80\% to the most central
0--5\% PbPb events, for the lowest \ptass\ interval of 1--1.5\GeVc. As \ptass\
increases, the centrality dependence of the near-side yield becomes less prominent.
An increase by a factor of only 1.3 is observed for the highest \ptass\ interval of
2.5--3\GeVc.
This is of particular interest because at RHIC energies for \ptass\ down to 2\GeVc
and similar \pttrg\ ranges and methodology, although for a lower density system
(AuAu at \rootsNN\ = 0.2\TeV), there is almost no centrality
dependence observed~\cite{star:2007pu}.
On both near and away sides, the yield in PbPb matches that
in pp for the most peripheral events.
On the away side, the yield in PbPb decreases with centrality, becoming negative
for the most central events. Variations in the event-mixing
procedure can cause large fluctuations but only at the very
edge of the acceptance around $|\Delta\eta| = 4.8$. However, the correlation
function is only studied up to $|\Delta\eta| < 4$ in this paper so these fluctuations do
not affect the results. The negative values of the yields in Fig.~\ref{fig:Correlation1D_4Panel_Yields_TrigPtBin_4} are
caused by a slightly concave structure on the away-side region
($1.18<|\Delta\phi|<\pi$), i.e. the yields near $|\Delta\eta|\approx 0$
are smaller than those at higher $|\Delta\eta|$. The effect
is more prominent for central PbPb events. However, this
concavity is seen only for $|\Delta\eta|<2$. Beyond that
region, the $\Delta\eta$ distribution is found
to be largely flat up to $|\Delta\eta|=4$. Similar behaviour was also observed at RHIC for AuAu collisions
at \rootsNN\ = 200\GeV~\cite{Agakishiev:2011pe}.
This deviation from pp may be related to the jet quenching phenomena,
which leads to a modification
in the back-to-back jet correlations in PbPb.
Any effect that modifies
the kinematics of dijet production could also
result in a modification of away-side distributions in $\Delta\eta$.
Additionally, any slight dependence of the flow effect on $\eta$ could also play a role.
More detailed theoretical models will be required to
fully understand the origin of this small effect.

\section{Fourier Decomposition Analysis of the PbPb Data}
\label{sec:PbPbFourierDecompositionAnalysis}

The first Fourier decomposition analysis
of long-range dihadron azimuthal correlations for PbPb collisions
at \rootsNN\ = 2.76\TeV was presented in Ref.~\cite{ref:HIN-11-001-PAS}.
This analysis was motivated by the goal of determining whether the long-range ridge effect
was caused by higher-order hydrodynamic flow harmonics induced by the initial
geometric fluctuations. This decomposition involves fitting the 1D $\Delta\phi$-projected
distribution for $2 < |\Delta\eta| < 4$ (to avoid the jet peak) with
a Fourier series given by
\begin{linenomath}
\begin{equation}
\label{fourier}
\frac{1}{N_\text{trig}}\frac{\d N^\text{pair}}{\d\Delta\phi} = \frac{N_\text{assoc}}{2\pi} \left\{ 1+\sum\limits_{n=1}^{N_\text{max}} 2V_{n\Delta} \cos (n\Delta\phi)\right\},
\end{equation}
\end{linenomath}
where $V_{n\Delta}$ are the Fourier coefficients and  $N_\text{assoc}$
represents the total number of hadron
pairs per trigger particle for the given $|\Delta\eta|$ range and
$(\pttrg, \ptass)$ bin. The first five Fourier terms ($N_\text{max}=5$) are included
in both the current fits and those in Ref.~\cite{ref:HIN-11-001-PAS}.
In this paper, the analysis of the Fourier
decomposition is extended to the full centrality range, and is performed as a function
of both \pttrg\ and \ptass.

The Fourier decomposition results have several systematic
uncertainties. Because the tracking-correction-related
systematic uncertainties 
only change the overall scale of the correlation functions, instead of
the shape, they have only a
$\pm$0.8\% uncertainty on the extracted Fourier coefficients ($V_{n\Delta}$),
largely independent of $n$ and collision centrality. In addition,
the results are insensitive to looser or tighter track
selections to within $\pm$0.5\%. By comparing the Fourier coefficients
derived for two different $z_\text{vtx}$ ranges, $|z_\text{vtx}| < 15\cm$
and $|z_\text{vtx}| < 5\cm$, the systematic uncertainties due to the
dependence on the vertex position are estimated to be less than
$\pm$0.5\%. Variations from the finite bin width of the
$\Delta\phi$ histograms contribute the largest systematic uncertainty to the analysis,
especially for the higher-order components, which are more sensitive to the
fine structure of the distributions. Reducing the binning of the
$\Delta\phi$ histograms by factors of 2, 4, and 8, the extracted
Fourier coefficients vary by $\pm$0.3--2.2\%.
The effect of including additional higher-order Fourier terms in the fit using
Eq.~\ref{fourier} results
in changes of at most $\pm$1.0\% (for $n=5$),
with Fourier terms up to $n=10$ included ($N_\text{max}=10$). The values of additional
higher-order Fourier terms included in the fit are all consistent with zero.

Table~\ref{tab:syst-table-fourier} summarises the different sources of uncertainty
for the first five Fourier coefficients. These uncertainties are added in quadrature
to obtain the total systematic uncertainties, also given in Table~\ref{tab:syst-table-fourier}.

\begin{table*}[ht]
\topcaption{\label{tab:syst-table-fourier} Systematic
uncertainties of the Fourier coefficients ($V_{n\Delta}$) for the first five terms.}

\begin{center}
\begin{tabular}{lc}
\hline
\hline
 Source                                          & Systematic uncertainty of $V_{n\Delta}$ (\%)\\
\hline
 Tracking efficiency                                                  & 0.8 \\
 Vertex dependence                                                    & 0.5 \\
 Track selection dependence                                           & 0.5 \\
 Finite bin width in $\Delta\phi$                                     & 0.3--2.2\\
 Number of terms included in the fit                                  & 0.0--1.0\\
\hline
 Total                                                                & 1.1--2.6\\
\hline
\hline
\end{tabular}
\end{center}
\end{table*}

\begin{figure*}[thbp]
  \begin{center}
    \includegraphics[width=\textwidth]{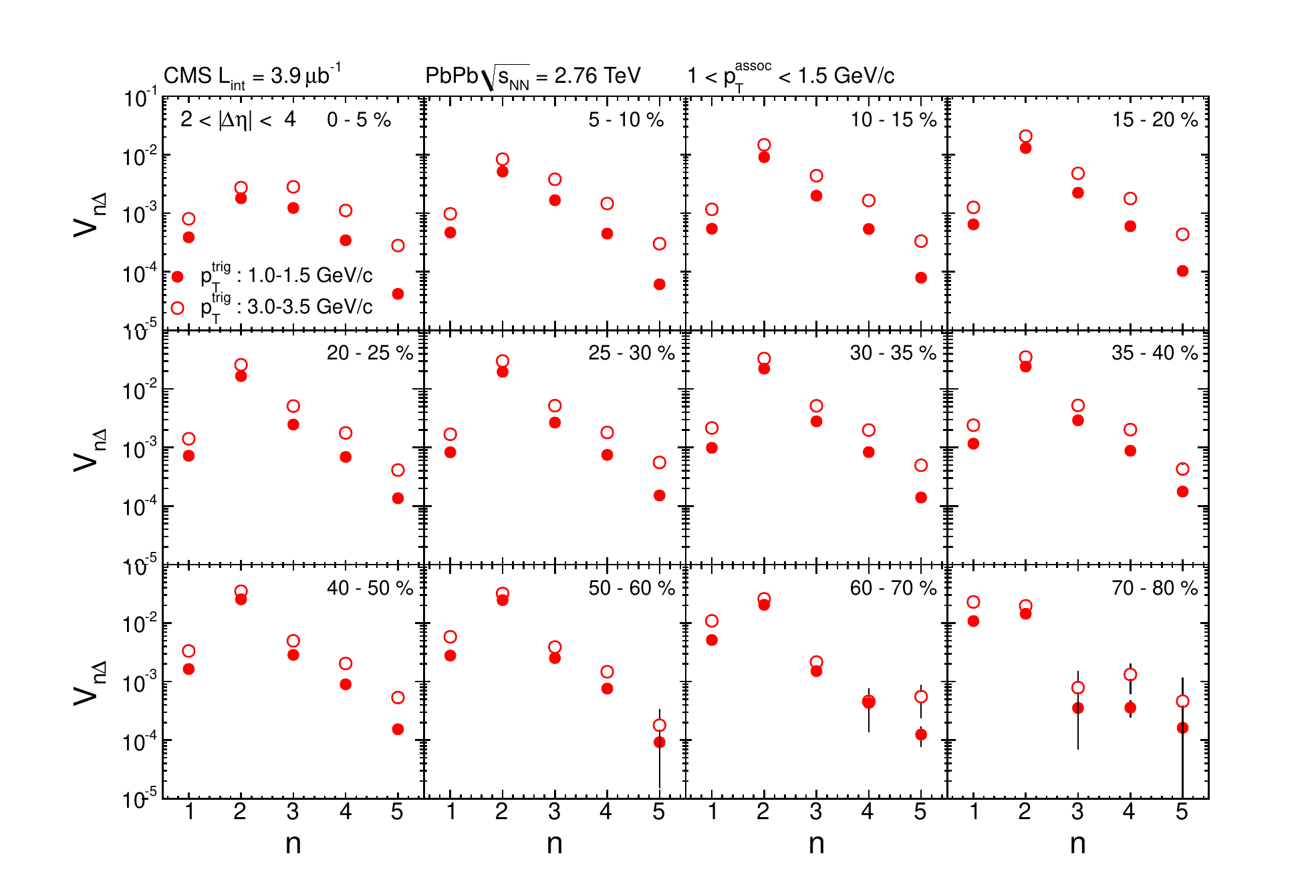}
    \caption{ Fourier coefficients $V_{1\Delta}$ through $V_{5\Delta}$, extracted from the
              long-range ($2<|\Delta\eta|<4$) azimuthal dihadron correlations,
              for $1 < \pttrg < 1.5$\GeVc (closed circles) and
              $3 < \pttrg < 3.5$\GeVc (open circles) with
              \ptass\ fixed at 1--1.5\GeVc,
              for twelve centrality intervals.
              Most of the statistical error bars are
              smaller than the marker size.
              The systematic uncertainties (not shown in the plots) are indicated in
              Table~\ref{tab:syst-table-fourier}.
    }
    \label{fig:allbigVn}
  \end{center}
\end{figure*}

The fitted Fourier coefficients ($V_{n\Delta}$)
up to $n=5$, for two representative low-\pttrg\ ranges
of $1 < \pttrg < 1.5$\GeVc and $3 < \pttrg < 3.5$\GeVc, with \ptass\
fixed at 1--1.5\GeVc, are presented in Fig.~\ref{fig:allbigVn}
for various centrality ranges.
The values of $V_{n\Delta}$ peak at $n=2$ and then drop dramatically
toward larger $n$ values at all centralities, although this behaviour is less
pronounced for the 0--5\% centrality bin. The error bars show the
statistical uncertainties only, while the systematic uncertainties are indicated
in Table~\ref{tab:syst-table-fourier}.

\subsection{Factorisation of Fourier Coefficients}
\label{subsec:factorization}
If the observed azimuthal dihadron correlations at large $\Delta\eta$
are driven only by the single-particle azimuthal anisotropy
with respect to a particular direction in the event,
the extracted Fourier coefficients ($V_{n\Delta}$) from long-range azimuthal
dihadron correlations can be factorised into a product of the single-particle azimuthal
anisotropy harmonics, $v_{n}$, via
\begin{linenomath}
\begin{equation}
\label{eq:factorization}
V_{n\Delta}(\pttrg, \ptass) = v_{n}(\pttrg) \times v_{n}(\ptass),
\end{equation}
\end{linenomath}
where $v_{n}(\pttrg)$ and $v_{n}(\ptass)$
are the harmonics for the trigger and associated
particles~\cite{Collaboration:2011by} averaged over all the events, respectively.
One source of $v_{n}$ is the collective-flow harmonics arising
from hydrodynamic expansion of the medium
(e.g., anisotropic elliptic flow contribution to $v_{2}$)~\cite{Voloshin:1994mz}, particularly
in the low-\pt\ regime where hadron production in heavy ion
collisions is thought to be mainly from the bulk medium~\cite{Voloshin:2008dg}.
In addition, for very high \pt\ particles that are predominantly
produced by the fragmentation of energetic jets,
$v_{n}$ could also be induced by the path-length dependence of the
jet-quenching effect inside the medium~\cite{Adare:2010sp,Bass:2008rv,
Peigne:2008wu,Gubser:2008as,Wicks:2005gt,Marquet:2009eq}. This path difference can lead
to a stronger suppression of the high-\pt\ hadron yield along the long axis
of the elliptically-shaped system than along its short axis, resulting in
an azimuthal anisotropy characterised by the $v_{2}$ harmonic. Both scenarios
satisfy the factorisation relation of Eq.~\ref{eq:factorization}. However,
note that the \pt\ dependent event-by-event fluctuations of $v_{n}$ could
break the factorisation in general, even though $v_{n}$ may be still related
to the single-particle azimuthal anisotropy. This possibility is not investigated in this paper.

This relation (Eq.~\ref{eq:factorization}) is a necessary ingredient
for the extraction of single-particle azimuthal anisotropy harmonics
using the dihadron correlation data, since a Fourier series can be used
to decompose any functional form by construction. The
relation can be tested by first assuming that factorisation is
valid for pairs including one particle in a fixed, low \ptass\
range, denoted by \ptlow, correlated with a second particle of any \pt.
The range of \ptlow\ is chosen to be 1--1.5\GeVc, where particle production
is expected to be predominantly driven by hydrodynamics. The value of $v_{n}(\ptlow)$
is first calculated as the square root of $V_{n\Delta}(\ptlow,\ptlow)$.
The $v_{n}(\pttrg)$ is then derived as

\begin{linenomath}
\begin{equation}
\label{eq:small_vn}
v_{n}(\pttrg) = \frac{V_{n\Delta}(\pttrg,\ptlow)}{v_{n}(\ptlow)}.
\end{equation}
\end{linenomath}

This is effectively equivalent to the two-particle cumulant method
of flow measurement~\cite{Borghini:2000sa,Borghini:2001vi}.
Next, the ratio of $V_{n\Delta}(\pttrg, \ptass)$ directly extracted as a function
of \pttrg\ and \ptass\ (left-hand side of Eq.~\ref{eq:factorization})
to the product of $v_{n}(\pttrg)$ and $v_{n}(\ptass)$
(right-hand side of Eq.~\ref{eq:factorization}) is calculated.
This ratio should be approximately unity if factorisation is also valid for higher
\pttrg\ and \ptass\ particles.

\begin{figure*}[thbp]
  \begin{center}
    \includegraphics[width=\textwidth]{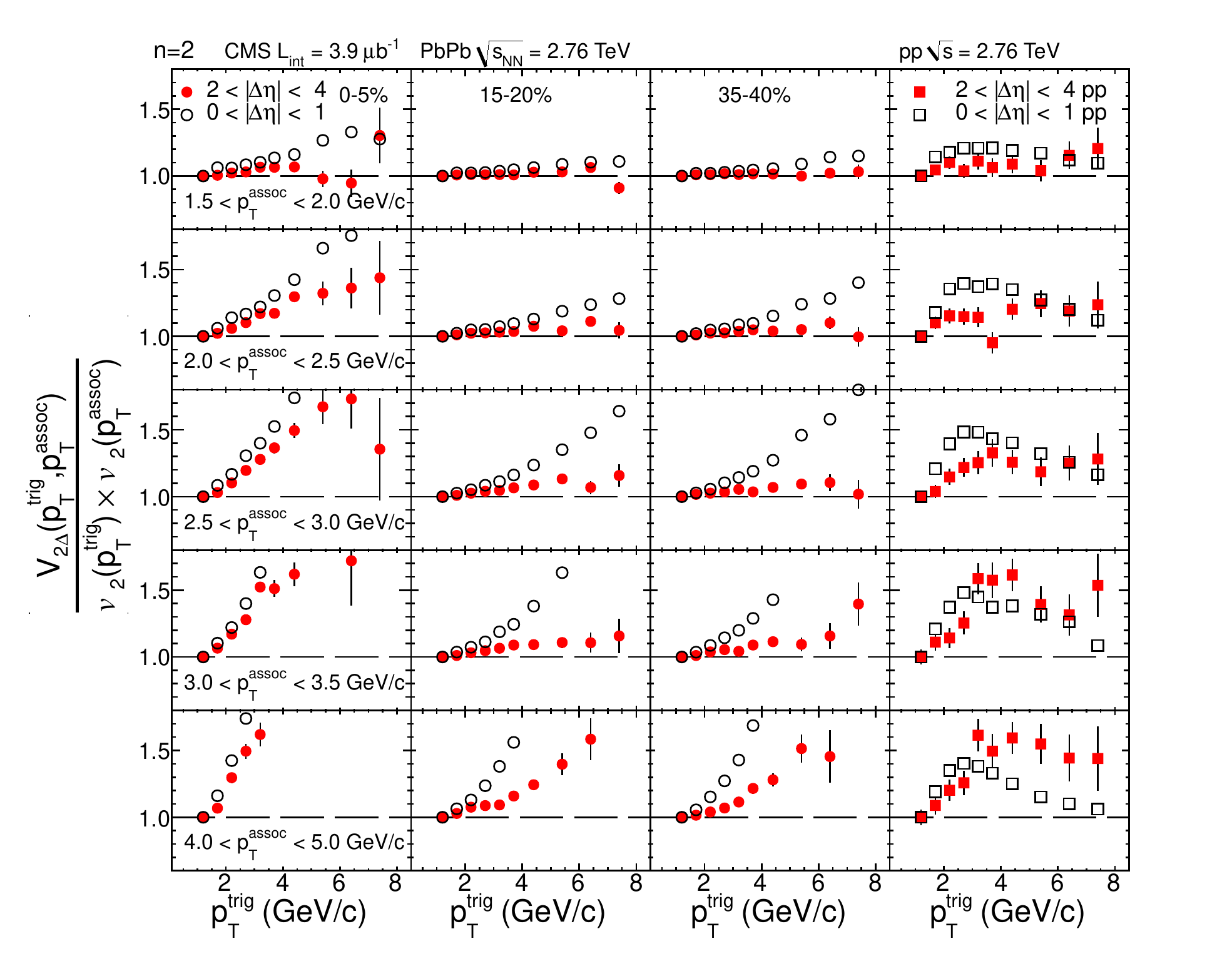}
    \caption{
       The ratios of $V_{2\Delta}(\pttrg, \ptass)$ to the product of
       $v_{2}(\pttrg)$ and $v_{2}(\ptass)$ for $n=2$ in the short-range
       ($0<|\Delta\eta|<1$, open circles) and long-range
       ($2<|\Delta\eta|<4$, closed circles) regions,
       where $v_{2}(\pt)$ is evaluated in a fixed \ptlow\ bin of 1--1.5\GeVc,
       for five intervals of \ptass\ and centralities of 0--5\%, 15--20\% and 35--40\%.
       The error bars correspond to statistical uncertainties only.
    }
    \label{fig:factorizationv2}
  \end{center}
\end{figure*}

\begin{figure*}[thbp]
  \begin{center}
    \includegraphics[width=\textwidth]{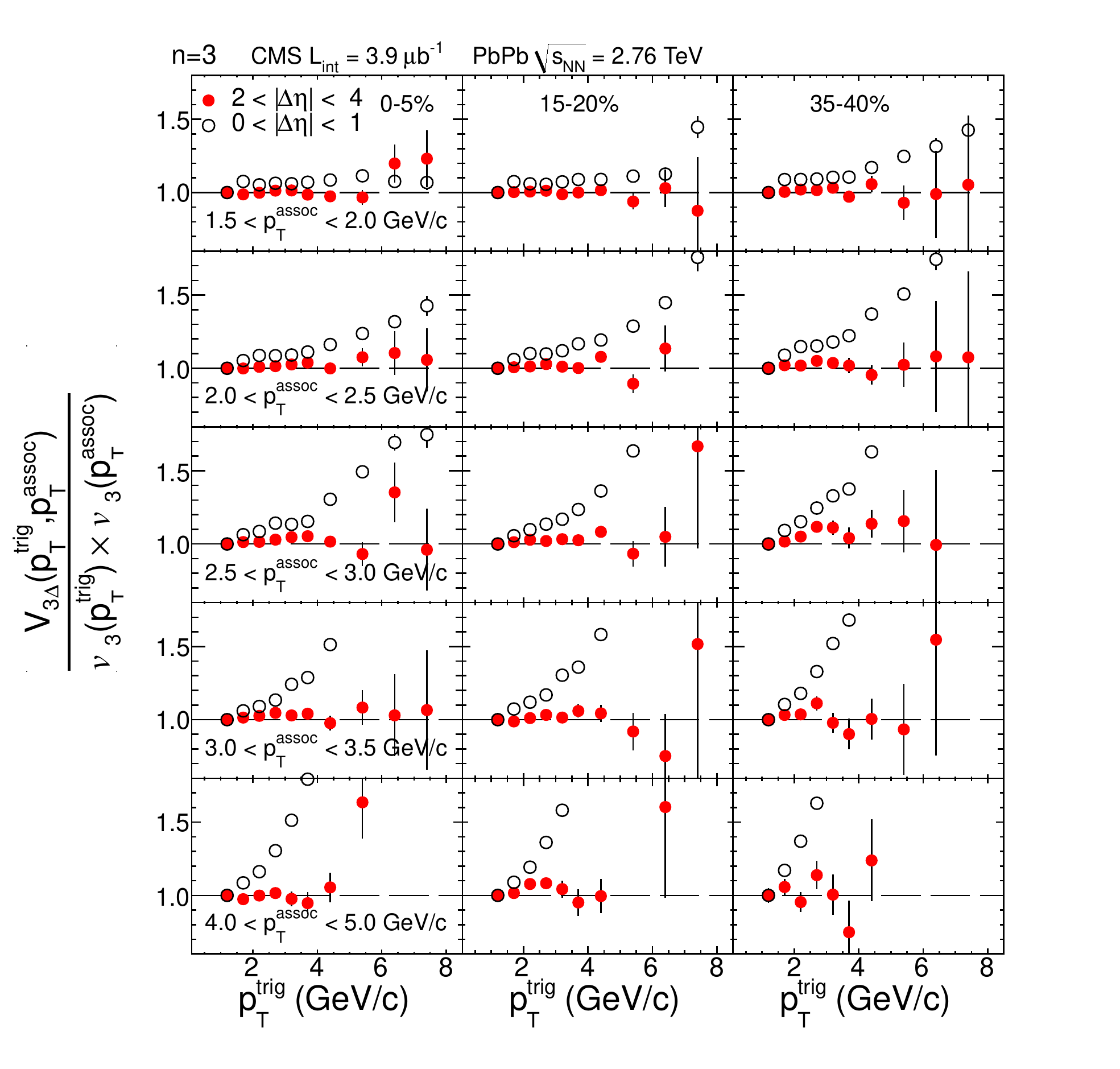}
    \caption{
      The ratios of $V_{3\Delta}(\pttrg, \ptass)$ to the product of
       $v_{3}(\pttrg)$ and $v_{3}(\ptass)$ for $n=3$ in the short-range
       ($0<|\Delta\eta|<1$, open circles) and long-range
       ($2<|\Delta\eta|<4$, closed circles) regions,
       where $v_{3}(\pt)$ is evaluated in a fixed \ptlow\ bin of 1--1.5\GeVc,
       for five intervals of \ptass\ and centralities of 0--5\%, 15--20\% and 35--40\%.
       The error bars correspond to statistical uncertainties only.
    }
    \label{fig:factorizationv3}
  \end{center}
\end{figure*}

\begin{figure*}[thbp]
  \begin{center}
    \includegraphics[width=\linewidth]{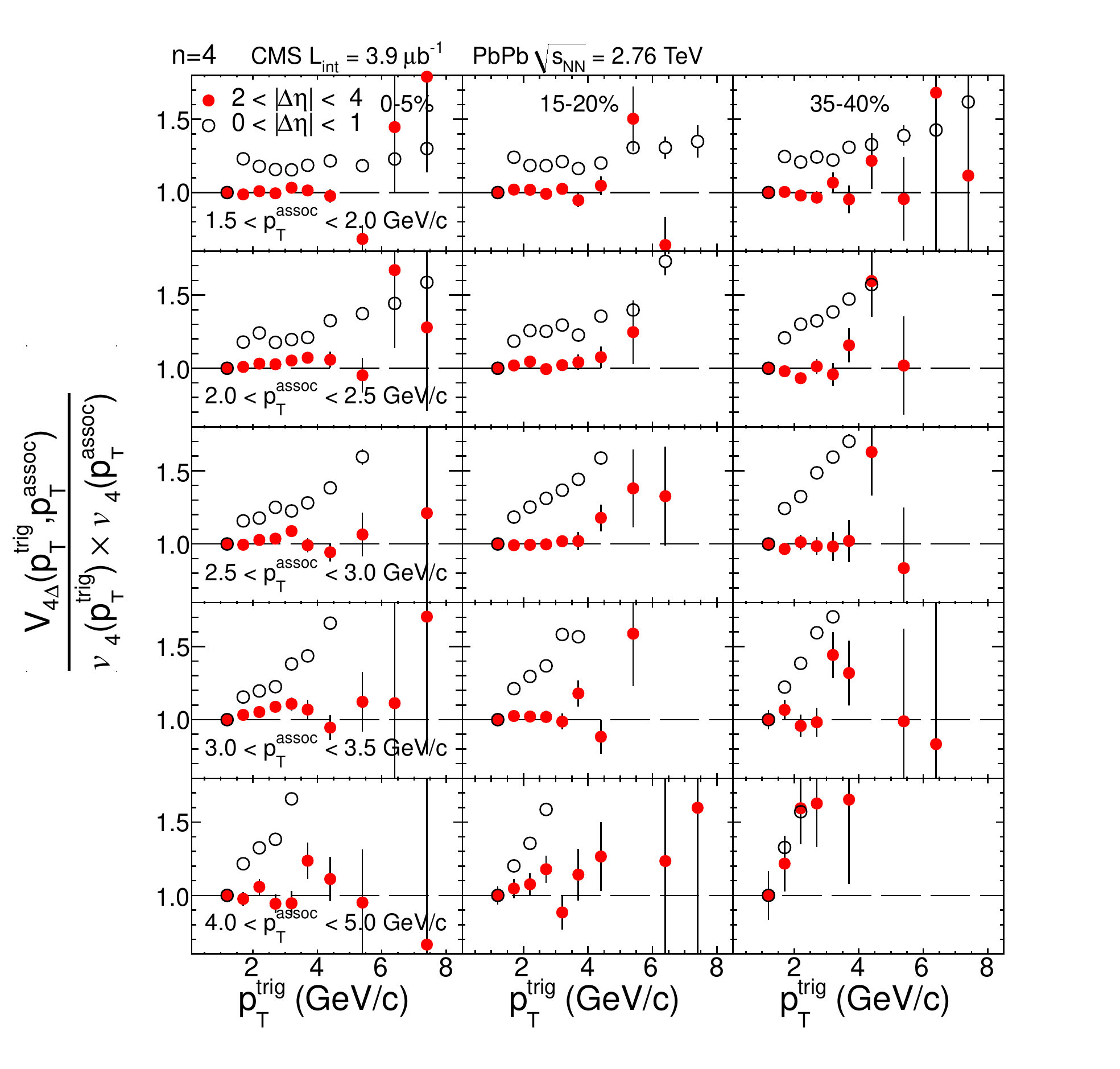}
    \caption{
       The ratios of $V_{4\Delta}(\pttrg, \ptass)$ to the product of
       $v_{4}(\pttrg)$ and $v_{4}(\ptass)$ for $n=4$ in the short-range
       ($0<|\Delta\eta|<1$, open circles) and long-range
       ($2<|\Delta\eta|<4$, closed circles) regions,
       where $v_{4}(\pt)$ is evaluated in a fixed \ptlow\ bin of 1--1.5\GeVc,
       for five intervals of \ptass\ and centralities of 0--5\%, 15--20\% and 35--40\%.
       The error bars correspond to statistical uncertainties only.
    }
    \label{fig:factorizationv4}
  \end{center}
\end{figure*}

Figures~\ref{fig:factorizationv2},~\ref{fig:factorizationv3}, and~\ref{fig:factorizationv4} show
the ratios for $n=2$, $3$, and $4$, over five \ptass\ ranges as a
function of \pttrg, for both short-range ($0<|\Delta\eta|<1$) and
long-range ($2<|\Delta\eta|<4$) regions. Three different centrality intervals 0--5\%,
15--20\% and 35--40\% are presented. The ratio for $n=2$ in pp data is also shown
in the last column of Fig.~\ref{fig:factorizationv2}. The first point of each panel
equals 1.0 by construction. The error bars correspond to
the statistical uncertainties. The total systematic uncertainties are estimated to be
1.5\% ($n=2$)--3.6\% ($n=5$), approximately $\sqrt{2}$ times the systematic uncertainties
of $V_{n\Delta}$ shown in Table~\ref{tab:syst-table-fourier}.

First of all, no evidence of factorisation is found in the pp data and
for the short-range region ($0<|\Delta\eta|<1$) in any of the centrality
ranges of the PbPb data, where dijet production is
expected to be the dominant source of correlations. In Ref.~\cite{Collaboration:2011by},
it has been shown that
$V_{n\Delta}$ factorises for jet-like correlations at very
high-\pt\ (e.g., $\pt>5\GeVc$) as the direction of the dijet
forms a special axis, to which produced particles are strongly correlated.
This is similar to the elliptic flow effect, where particles
are preferentially produced along the short axis of the
elliptically-shaped overlapping region. However, the lack of factorisation
in pp and the short-range region of PbPb
observed for the \pt\ range of $1<\ptass<3.5$\GeVc primarily investigated
in this paper may suggest a complicated interplay of different
particle production mechanisms between low-\pt\ (hydrodynamic flow
for PbPb and underlying event for pp) and high-\pt\ (dijet production)
particles. 

In contrast, the long-range region ($2<|\Delta\eta|<4$) for
mid-peripheral 15--20\% and 35--40\% events does show evidence
of factorisation for $V_{2\Delta}$ to $V_{4\Delta}$
with \ptass\ up to 3--3.5\GeVc and \pttrg\ up to approximately 8\GeVc.
The data are also consistent with factorisation for even
higher \pttrg\ ($>8\GeVc$) combined with low \ptass,
but the current event sample is not large enough to draw a firm conclusion.
Note that $V_{n\Delta}$ varies by almost 60\% in the \pttrg\ range from
1 to 3.5\GeVc as shown in Fig.~\ref{fig:allbigVn}, whereas
factorisation is found to hold to better than 5\%. This suggests a potential
connection between the extracted Fourier coefficients from long-range
dihadron correlations and the single-particle azimuthal anisotropy
harmonics. For the most central 0--5\% collisions,
the ratio 
for $n=2$ deviates significantly from unity,
while $V_{3\Delta}$ and $V_{4\Delta}$ still show a similar level of
factorisation to that of the 15--20\% and 35--40\% mid-peripheral data.
This may indicate the existence of other sources of long-range
correlations for the most central collisions that violate
the factorisation relation.
The breakdown of factorisation for $\ptass>4$\GeVc
in the long-range region is likely due to dijet correlations.
Higher-order Fourier coefficients
and a wider \pt\ range can be investigated once larger samples
of PbPb data are collected. Factorisation of $V_{1\Delta}$ is not discussed
in this paper as it contains an additional negative contribution
from momentum conservation~\cite{Luzum:2010fb}, which is not related to the
collective flow effect and needs to be accounted for in further studies.

\subsection{Elliptic and Higher-Order Single-Particle Azimuthal Anisotropy Harmonics}

\begin{figure*}[thbp]
  \begin{center}
    \includegraphics[width=\linewidth]{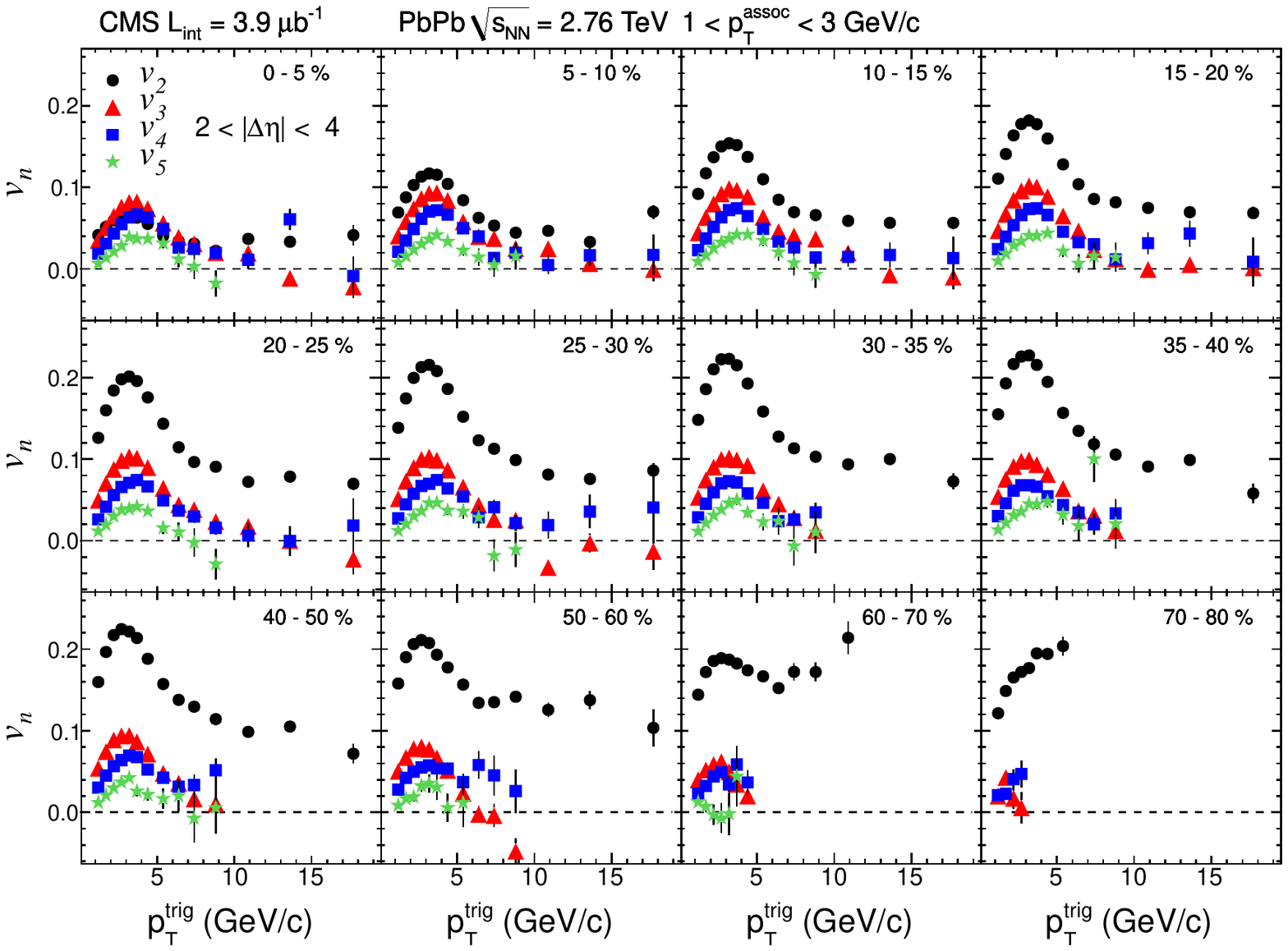}
    \caption{ The single-particle azimuthal anisotropy harmonics
              $v_{2}$--$v_{5}$ extracted from the long-range ($2<|\Delta\eta|<4$)
              azimuthal dihadron correlations as a function of \pttrg, combined
              with $1<\ptass<3\GeVc$, for twelve centrality intervals in PbPb collisions
              at \rootsNN\ = 2.76\TeV. Most of the statistical error bars are
              smaller than the marker size. The systematic uncertainties
              (not shown in the plots) are indicated in
              Table~\ref{tab:syst-table-fourier}.
    }
    \label{fig:allsmallvn}
  \end{center}
\end{figure*}

\begin{figure*}[thbp]
  \begin{center}
    \includegraphics[width=\linewidth]{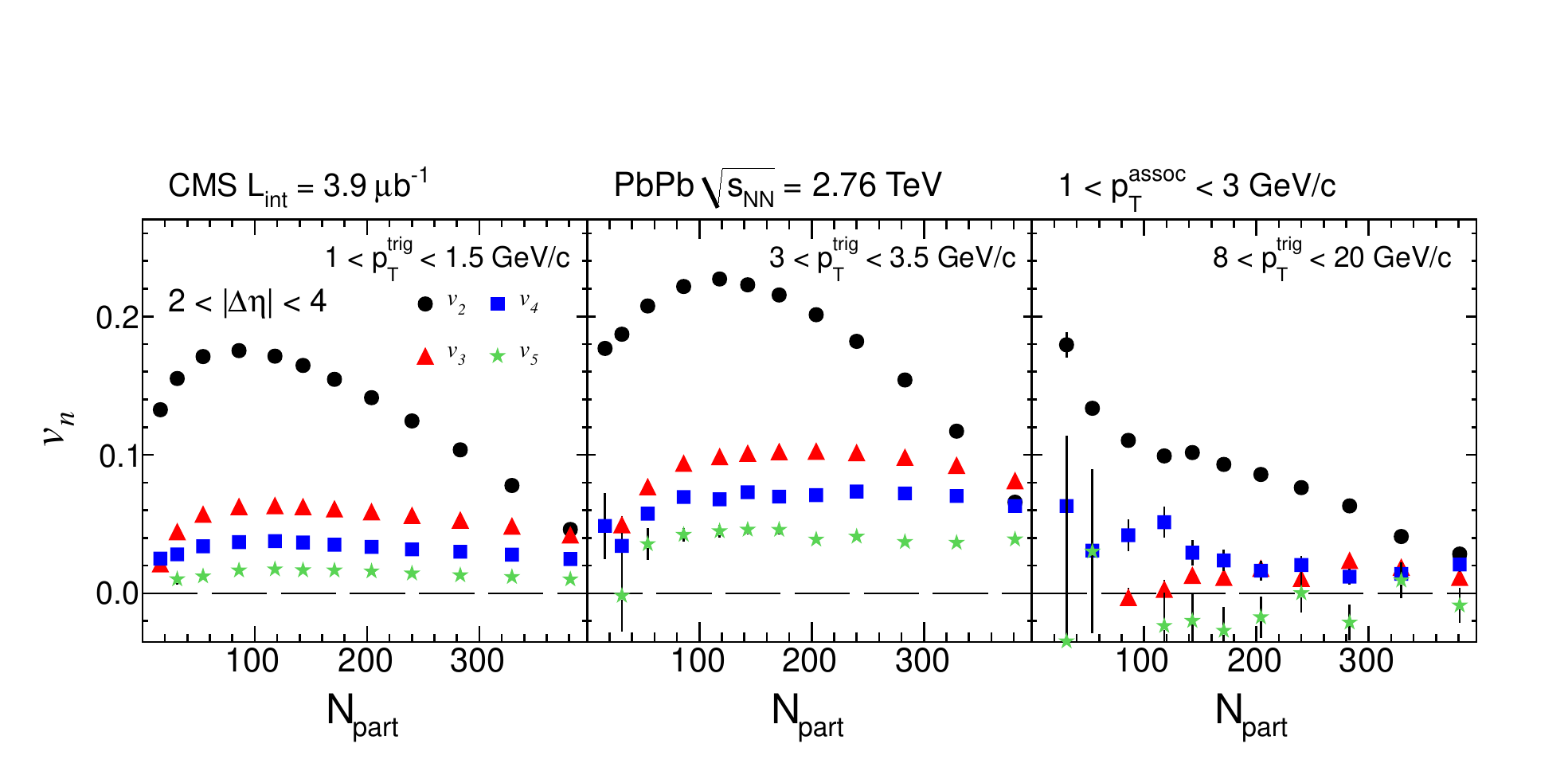}
    \caption{
              The single-particle azimuthal anisotropy harmonics,
              $v_{2}$--$v_{5}$, extracted from the long-range ($2<|\Delta\eta|<4$)
              azimuthal dihadron correlations as a function of \npart\ in PbPb collisions
              at \rootsNN\ = 2.76\TeV for $1<\ptass<3\GeVc$ in three
              \pttrg\ ranges of 1--1.5, 3--3.5 and 8--20\GeVc.
              Most of the statistical error bars are
              smaller than the marker size.
              The systematic uncertainties (not shown in the plots) are indicated in
              Table~\ref{tab:syst-table-fourier}.
    }
    \label{fig:vn_npart}
  \end{center}
\end{figure*}

As discussed in Section~\ref{subsec:factorization}, except
for $v_{2}$ in the very central PbPb events, the factorisation relation given by
Eq.~\ref{eq:factorization} for long-range ($2<|\Delta\eta|<4$)
azimuthal dihadron correlations is found to be valid for sufficiently low
\ptass, combined with low \pttrg, and possibly
high \pttrg\ as well. Therefore, the single-particle azimuthal
anisotropy harmonics $v_{n}(\pttrg)$
can be extracted using Eq.~\ref{eq:small_vn} with $1<\ptass<3\GeVc$. Values are found
for centralities ranging from 0--5\% to 70--80\%,
and presented in Fig.~\ref{fig:allsmallvn}. The 1--3\GeVc\ \ptass\ range
is chosen in order to reduce the statistical uncertainty
by utilising as many associated particles
as possible over the \ptass\ range where factorisation is valid.
Data for the most central and most peripheral events are included for completeness, although
the results for $v_{2}$ in those events are clearly demonstrated by
Fig.~\ref{fig:factorizationv2} to be more complicated in nature.
The value of $v_{2}$ is extracted up to \pttrg\ $\sim 20\GeVc$
for all but the 2 most peripheral centralities, whereas higher-order $v_{n}$ are
truncated at \pttrg\ $\sim 10\GeVc$ or less for the peripheral
data due to statistical limitations. As mentioned previously,
factorisation is not demonstrated conclusively at very high \pttrg.
For the most central 0--5\% events, all the harmonics are
of similar magnitude across the entire \pttrg\ range. The \pt\
dependence of all $v_{n}$
shows the same trend of a fast rise to a maximum around
\pt\ $\approx$ 3 \GeVc, followed by a slower fall, independent of
centrality up to 50--60\%. The magnitude of $v_{2}$ increases
when moving away from the most central events. At very high \pttrg,
sizeable $v_{2}$ signals are observed, which exhibit
an almost flat \pt\ dependence from 10 to 20\GeVc for most of
the centrality ranges. This is not the case for the higher-order harmonics.
In order to explicitly investigate the centrality
dependence of the harmonics, the extracted $v_{2}$ through $v_{5}$
are also shown in Fig.~\ref{fig:vn_npart} as a function
of \npart, for representative low (1--1.5\GeVc), intermediate
(3--3.5\GeVc), and high (8--20\GeVc) \pttrg\ ranges.

A strong centrality dependence of $v_{2}$ is
observed in Figs.~\ref{fig:allsmallvn} and~\ref{fig:vn_npart}
for all \pttrg\ ranges, while the higher-order harmonics
$v_{3}$--$v_{5}$ do not vary significantly over the range of
\npart. This behaviour is expected in
the context of both the hydrodynamic flow phenomena for lower-\pt\ particles~\cite{Voloshin:2008dg}
and the path-length dependence of the parton energy-loss scenario for high-\pt\ particles~\cite{Adare:2010sp}.
The $v_{2}$ harmonics are sensitive to the eccentricity of
the almond-shaped initial collision region
that becomes larger for the peripheral events, whereas the
higher-order harmonics are driven by fluctuations of the
initial geometry that have little dependence on the collision
centrality~\cite{Alver:2008zza}.
In the most peripheral events, the \pt\ dependence of $v_{2}$ is
found to be very different from that in the central events,
as shown in Fig.~\ref{fig:allsmallvn}. In the high-\pt\ interval
8--20\GeVc (third panel of Fig.~\ref{fig:vn_npart}), $v_{2}$
increases rapidly at low \npart\ (very peripheral). A possible explanation
is the presence of non-flow effects due to back-to-back jets.
Based on the factorised Fourier coefficients from long-range
dihadron correlations, the single-particle azimuthal anisotropy
harmonics extracted over a wide range of \pt\ and centrality
allow a detailed comparison to theoretical calculations of the hydrodynamics
and path-length dependence of in-medium parton energy loss.

\section{Summary}
\label{sec:conclusion}
The previous CMS analysis of angular correlations between charged
particles has been expanded to cover a wide centrality range of PbPb collisions
at \rootsNN\ = 2.76\TeV. As was seen previously for central PbPb collisions,
the associated yields differ significantly from those observed in pp interactions.
Correlations with both small ($0 < |\Delta\eta| < 1$) and large
($2 < |\Delta\eta| < 4$) relative pseudorapidities were again studied
as a function of the transverse momentum of the trigger and associated
particle \pt. The integrated yield of the near-side region shows an
increasing enhancement towards more central PbPb collisions, especially
for low-\pt\ associated particles.

To further characterise the dependence of the correlations on
relative azimuthal angle, a Fourier decomposition of the distributions
projected onto $\Delta\phi$ was performed, as a function of both
centrality and particle \pt. Evidence of a factorisation relation
was observed between the Fourier coefficients ($V_{n\Delta}$)
from dihadron correlations and single-particle azimuthal
anisotropy harmonics ($v_{n}$). This holds for
\ptass\ $\lesssim$ 3--3.5\GeVc over the \pttrg\ range up
to at least 8\GeVc in central and mid-peripheral PbPb collisions,
except for $v_{2}$ in the most central events. The observed factorisation
is absent in pp and very peripheral PbPb data, indicating
a strong connection of the observed long-range azimuthal dihadron
correlations to the single-particle azimuthal anisotropy
in heavy ion collisions, such as the one driven by the hydrodynamic expansion of the
system. The single-particle azimuthal anisotropy harmonics
$v_{2}$ through $v_{5}$ were extracted over a wide range in both \pt\ and collision centrality,
profiting from the broad solid-angle coverage of the CMS detector.
The comprehensive correlation data presented in this paper
are very useful for studies of the path-length dependence of
in-medium parton energy-loss, and provide
valuable inputs to a variety of theoretical models, including
hydrodynamic calculations of higher-order Fourier components.

\section*{Acknowledgments}
We wish to congratulate our colleagues in the CERN accelerator departments for the excellent performance of the LHC machine. We thank the technical and administrative staff at CERN and other CMS institutes, and acknowledge support from: FMSR (Austria); FNRS and FWO (Belgium); CNPq, CAPES, FAPERJ, and FAPESP (Brazil); MES (Bulgaria); CERN; CAS, MoST, and NSFC (China); COLCIENCIAS (Colombia); MSES (Croatia); RPF (Cyprus); Academy of Sciences and NICPB (Estonia); Academy of Finland, MEC, and HIP (Finland); CEA and CNRS/IN2P3 (France); BMBF, DFG, and HGF (Germany); GSRT (Greece); OTKA and NKTH (Hungary); DAE and DST (India); IPM (Iran); SFI (Ireland); INFN (Italy); NRF and WCU (Korea); LAS (Lithuania); CINVESTAV, CONACYT, SEP, and UASLP-FAI (Mexico); MSI (New Zealand); PAEC (Pakistan); SCSR (Poland); FCT (Portugal); JINR (Armenia, Belarus, Georgia, Ukraine, Uzbekistan); MON, RosAtom, RAS and RFBR (Russia); MSTD (Serbia); MICINN and CPAN (Spain); Swiss Funding Agencies (Switzerland); NSC (Taipei); TUBITAK and TAEK (Turkey); STFC (United Kingdom); DOE and NSF (USA).

Individuals have received support from the Marie-Curie programme and the European Research Council (European Union); the Leventis Foundation; the A. P. Sloan Foundation; the Alexander von Humboldt Foundation; the Belgian Federal Science Policy Office; the Fonds pour la Formation \`a la Recherche dans l'Industrie et dans l'Agriculture (FRIA-Belgium); the Agentschap voor Innovatie door Wetenschap en Technologie (IWT-Belgium); and the Council of Science and Industrial Research, India.

\bibliography{auto_generated}   

\providecommand{\href}[2]{#2}\begingroup\raggedright\begin{thebibliography}{10}%
\makeatletter
\providecommand{\hrefCMSnoop }[0]{\@secondoftwo}%
\makeatother
\providecommand{\doi}{\texttt{doi:}\begingroup \urlstyle{tt}\Url}

\bibitem{ref:HIN-11-001-PAS}
\hrefCMSnoop {} {{ CMS} Collaboration, ``{Long-range and short-range dihadron
  angular correlations in central Pb Pb collisions at \rootsNN\ = 2.76 TeV}'',}
  \textit{ JHEP} \textbf{ 07} (2011) 076,
  \href{http://dx.doi.org/10.1007/JHEP07(2011)076}{\doi{10.1007/JHEP07(2011)076}},
\href{http://www.arXiv.org/abs/1105.2438}{\texttt{ arXiv:1105.2438}}.

\bibitem{Collaboration:2011by}
\hrefCMSnoop {} {{ALICE Collaboration}, ``{Harmonic decomposition of
  two-particle angular correlations in Pb--Pb collisions at \rootsNN\ = 2.76
  TeV}'',} (2011).
\href{http://www.arXiv.org/abs/1109.2501}{\texttt{ arXiv:1109.2501}}.

\bibitem{Adler:2002tq}
\hrefCMSnoop {} {{ STAR} Collaboration, ``{Disappearance of back-to-back high
  $p_{T}$ hadron correlations in central Au+Au collisions at \rootsNN\ = 200
  GeV}'',} \textit{ Phys. Rev. Lett.} \textbf{ 90} (2003) 082302,
  \href{http://dx.doi.org/10.1103/PhysRevLett.90.082302}{\doi{10.1103/PhysRevLett.90.082302}},
\href{http://www.arXiv.org/abs/nucl-ex/0210033}{\texttt{
  arXiv:nucl-ex/0210033}}.

\bibitem{star:2009qa}
\hrefCMSnoop {} {{ STAR} Collaboration, ``{Long range rapidity correlations and
  jet production in high energy nuclear collisions}'',} \textit{ Phys. Rev. C}
  \textbf{ 80} (2009) 064912,
  \href{http://dx.doi.org/10.1103/PhysRevC.80.064912}{\doi{10.1103/PhysRevC.80.064912}},
\href{http://www.arXiv.org/abs/0909.0191}{\texttt{ arXiv:0909.0191}}.

\bibitem{star:2010ridge}
\hrefCMSnoop {} {{ STAR} Collaboration, ``{System size dependence of associated
  yields in hadron-triggered jets}'',} \textit{ Phys. Lett. B} \textbf{ 683}
  (2010) 123,
  \href{http://dx.doi.org/10.1103/PhysRevC.80.064912}{\doi{10.1103/PhysRevC.80.064912}},
\href{http://www.arXiv.org/abs/0909.0191}{\texttt{ arXiv:0909.0191}}.

\bibitem{Adams:2005ph}
\hrefCMSnoop {} {{ STAR} Collaboration, ``{Distributions of charged hadrons
  associated with high transverse momentum particles in pp and Au+Au collisions
  at \rootsNN\ = 200 GeV}'',} \textit{ Phys. Rev. Lett.} \textbf{ 95} (2005)
  152301,
  \href{http://dx.doi.org/10.1103/PhysRevLett.95.152301}{\doi{10.1103/PhysRevLett.95.152301}},
\href{http://www.arXiv.org/abs/nucl-ex/0501016}{\texttt{
  arXiv:nucl-ex/0501016}}.

\bibitem{Adare:2006nr}
\hrefCMSnoop {} {{ PHENIX} Collaboration, ``{System size and energy dependence
  of jet-induced hadron pair correlation shapes in Cu+Cu and Au+Au collisions
  at \rootsNN\ = 200 and 62.4 GeV}'',} \textit{ Phys. Rev. Lett.} \textbf{ 98}
  (2007) 232302,
  \href{http://dx.doi.org/10.1103/PhysRevLett.98.232302}{\doi{10.1103/PhysRevLett.98.232302}},
\href{http://www.arXiv.org/abs/nucl-ex/0611019}{\texttt{
  arXiv:nucl-ex/0611019}}.

\bibitem{phenix:2008cqb}
\hrefCMSnoop {} {{ PHENIX} Collaboration, ``{Dihadron azimuthal correlations in
  Au+Au collisions at \rootsNN\ = 200 GeV}'',} \textit{ Phys. Rev. C} \textbf{
  78} (2008) 014901,
  \href{http://dx.doi.org/10.1103/PhysRevC.78.014901}{\doi{10.1103/PhysRevC.78.014901}},
\href{http://www.arXiv.org/abs/0801.4545}{\texttt{ arXiv:0801.4545}}.

\bibitem{Alver:2009id}
\hrefCMSnoop {} {{ PHOBOS} Collaboration, ``{High transverse Momentum Triggered
  Correlations over a Large Pseudorapidity Acceptance in Au+Au Collisions at
  $\sqrt{s_{_{NN}}}$ = 200~GeV}'',} \textit{ Phys. Rev. Lett.} \textbf{ 104}
  (2010) 062301,
  \href{http://dx.doi.org/10.1103/PhysRevLett.104.062301}{\doi{10.1103/PhysRevLett.104.062301}},
\href{http://www.arXiv.org/abs/0903.2811}{\texttt{ arXiv:0903.2811}}.

\bibitem{Alver:2008gk}
\hrefCMSnoop {} {{ PHOBOS} Collaboration, ``System size dependence of cluster
  properties from two- particle angular correlations in {Cu+Cu} and {Au+Au}
  collisions at $\sqrt{s_{_{NN}}}$ = 200~{GeV}'',} \textit{ Phys. Rev. C}
  \textbf{ 81} (2010) 024904,
  \href{http://dx.doi.org/10.1103/PhysRevC.81.024904}{\doi{10.1103/PhysRevC.81.024904}},
\href{http://www.arXiv.org/abs/0812.1172}{\texttt{ arXiv:0812.1172}}.

\bibitem{Aamodt:2010jd}
\hrefCMSnoop {} {{ ALICE} Collaboration, ``Suppression of charged particle
  production at large Transverse Momentum in Central Pb--Pb Collisions at
  $\sqrt{s_{_{NN}}} = 2.76$ {TeV}'',} \textit{ Phys. Lett. B} \textbf{ 696}
  (2011) 30,
  \href{http://dx.doi.org/10.1016/j.physletb.2010.12.020}{\doi{10.1016/j.physletb.2010.12.020}},
\href{http://www.arXiv.org/abs/1012.1004}{\texttt{ arXiv:1012.1004}}.

\bibitem{Aad:2010bu}
\hrefCMSnoop {} {{ ATLAS} Collaboration, ``{Observation of a
  Centrality-Dependent Dijet Asymmetry in Lead-Lead Collisions at \rootsNN\ =
  2.76\TeV with the ATLAS Detector at the LHC}'',} \textit{ Phys. Rev. Lett.}
  \textbf{ 105} (2010) 252303,
  \href{http://dx.doi.org/10.1103/PhysRevLett.105.252303}{\doi{10.1103/PhysRevLett.105.252303}},
\href{http://www.arXiv.org/abs/1011.6182}{\texttt{ arXiv:1011.6182}}.

\bibitem{HIN-10-004}
\hrefCMSnoop {} {{ CMS} Collaboration, ``{Observation and studies of jet
  quenching in PbPb collisions at nucleon-nucleon center-of-mass energy = 2.76
  TeV}'',} \textit{ Phys. Rev. C} \textbf{ 84} (2011) 024906,
  \href{http://dx.doi.org/10.1103/PhysRevC.84.024906}{\doi{10.1103/PhysRevC.84.024906}},
  \href{http://www.arXiv.org/abs/1102.1957}{\texttt{ arXiv:1102.1957}}.

\bibitem{Armesto:2004pt}
\hrefCMSnoop {} {N.~Armesto, C.~A. Salgado, and U.~A. Wiedemann, ``{Measuring
  the collective flow with jets}'',} \textit{ Phys. Rev. Lett.} \textbf{ 93}
  (2004) 242301,
  \href{http://dx.doi.org/10.1103/PhysRevLett.93.242301}{\doi{10.1103/PhysRevLett.93.242301}},
\href{http://www.arXiv.org/abs/hep-ph/0405301}{\texttt{ arXiv:hep-ph/0405301}}.

\bibitem{Majumder:2006wi}
\hrefCMSnoop {} {A.~Majumder, B.~Muller, and S.~A. Bass, ``{Longitudinal
  Broadening of Quenched Jets in Turbulent Color Fields}'',} \textit{ Phys.
  Rev. Lett.} \textbf{ 99} (2007) 042301,
  \href{http://dx.doi.org/10.1103/PhysRevLett.99.042301}{\doi{10.1103/PhysRevLett.99.042301}},
\href{http://www.arXiv.org/abs/hep-ph/0611135}{\texttt{ arXiv:hep-ph/0611135}}.

\bibitem{Chiu:2005ad}
\hrefCMSnoop {} {C.~B. Chiu and R.~C. Hwa, ``{Pedestal and peak structure in
  jet correlation}'',} \textit{ Phys. Rev. C} \textbf{ 72} (2005) 034903,
  \href{http://dx.doi.org/10.1103/PhysRevC.72.034903}{\doi{10.1103/PhysRevC.72.034903}},
\href{http://www.arXiv.org/abs/nucl-th/0505014}{\texttt{
  arXiv:nucl-th/0505014}}.

\bibitem{Wong:2008yh}
\hrefCMSnoop {} {C.-Y. Wong, ``Momentum kick model description of the near-side
  ridge and jet quenching'',} \textit{ Phys. Rev. C} \textbf{ 78} (2008)
  064905,
  \href{http://dx.doi.org/10.1103/PhysRevC.78.064905}{\doi{10.1103/PhysRevC.78.064905}},
\href{http://www.arXiv.org/abs/0806.2154}{\texttt{ arXiv:0806.2154}}.

\bibitem{Romatschke:2006bb}
\hrefCMSnoop {} {P.~Romatschke, ``{Momentum broadening in an anisotropic
  plasma}'',} \textit{ Phys. Rev. C} \textbf{ 75} (2007) 014901,
  \href{http://dx.doi.org/10.1103/PhysRevC.75.014901}{\doi{10.1103/PhysRevC.75.014901}},
\href{http://www.arXiv.org/abs/hep-ph/0607327}{\texttt{ arXiv:hep-ph/0607327}}.

\bibitem{Shuryak:2007fu}
\hrefCMSnoop {} {E.~V. Shuryak, ``On the origin of the ``ridge" phenomenon
  induced by jets in heavy ion collisions'',} \textit{ Phys. Rev. C} \textbf{
  76} (2007) 047901,
  \href{http://dx.doi.org/10.1103/PhysRevC.76.047901}{\doi{10.1103/PhysRevC.76.047901}},
\href{http://www.arXiv.org/abs/0706.3531}{\texttt{ arXiv:0706.3531}}.

\bibitem{Voloshin:2004th}
\hrefCMSnoop {} {S.~A. Voloshin, ``{Two particle rapidity, transverse momentum,
  and azimuthal correlations in relativistic nuclear collisions and transverse
  radial expansion}'',} \textit{ Nucl. Phys. A} \textbf{ 749} (2005) 287,
  \href{http://dx.doi.org/10.1016/j.nuclphysa.2004.12.053}{\doi{10.1016/j.nuclphysa.2004.12.053}},
\href{http://www.arXiv.org/abs/nucl-th/0410024}{\texttt{
  arXiv:nucl-th/0410024}}.

\bibitem{Mishra:2007tw}
\hrefCMSnoop {} {A.~P. Mishra {et~al.}, ``{Super-horizon fluctuations and
  acoustic oscillations in relativistic heavy-ion collisions}'',} \textit{
  Phys. Rev. C} \textbf{ 77} (2008) 064902,
  \href{http://dx.doi.org/10.1103/PhysRevC.77.064902}{\doi{10.1103/PhysRevC.77.064902}},
\href{http://www.arXiv.org/abs/0711.1323}{\texttt{ arXiv:0711.1323}}.

\bibitem{Takahashi:2009na}
\hrefCMSnoop {} {J.~Takahashi {et~al.}, ``{Topology studies of hydrodynamics
  using two particle correlation analysis}'',} \textit{ Phys. Rev. Lett.}
  \textbf{ 103} (2009) 242301,
  \href{http://dx.doi.org/10.1103/PhysRevLett.103.242301}{\doi{10.1103/PhysRevLett.103.242301}},
\href{http://www.arXiv.org/abs/0902.4870}{\texttt{ arXiv:0902.4870}}.

\bibitem{Alver:2010gr}
\hrefCMSnoop {} {B.~Alver and G.~Roland, ``{Collision-geometry fluctuations and
  triangular flow in heavy-ion collisions}'',} \textit{ Phys. Rev. C} \textbf{
  81} (2010) 054905,
  \href{http://dx.doi.org/10.1103/PhysRevC.81.054905}{\doi{10.1103/PhysRevC.81.054905}},
\href{http://www.arXiv.org/abs/1003.0194}{\texttt{ arXiv:1003.0194}}.

\bibitem{Alver:2010dn}
\hrefCMSnoop {} {B.~H. Alver {et~al.}, ``{Triangular flow in hydrodynamics and
  transport theory}'',} \textit{ Phys. Rev. C} \textbf{ 82} (2010) 034913,
  \href{http://dx.doi.org/10.1103/PhysRevC.82.034913}{\doi{10.1103/PhysRevC.82.034913}},
\href{http://www.arXiv.org/abs/1007.5469}{\texttt{ arXiv:1007.5469}}.

\bibitem{Schenke:2010rr}
\hrefCMSnoop {} {B.~Schenke, S.~Jeon, and C.~Gale, ``{Elliptic and triangular
  flow in event-by-event $D=(3+1)$ viscous hydrodynamics}'',} \textit{ Phys.
  Rev. Lett.} \textbf{ 106} (2011) 042301,
  \href{http://dx.doi.org/10.1103/PhysRevLett.106.042301}{\doi{10.1103/PhysRevLett.106.042301}},
\href{http://www.arXiv.org/abs/1009.3244}{\texttt{ arXiv:1009.3244}}.

\bibitem{Petersen:2010cw}
\hrefCMSnoop {} {H.~Petersen {et~al.}, ``{Triangular flow in event-by-event
  ideal hydrodynamics in Au+Au collisions at \rootsNN\ = 200 GeV}'',} \textit{
  Phys. Rev. C} \textbf{ 82} (2010) 041901,
  \href{http://dx.doi.org/10.1103/PhysRevC.82.041901}{\doi{10.1103/PhysRevC.82.041901}},
\href{http://www.arXiv.org/abs/1008.0625}{\texttt{ arXiv:1008.0625}}.

\bibitem{Xu:2010du}
\hrefCMSnoop {} {J.~Xu and C.~M. Ko, ``Effects of triangular flow on di-hadron
  azimuthal correlations in relativistic heavy ion collisions'',} \textit{
  Phys. Rev. C} \textbf{ 83} (2011) 021903,
  \href{http://dx.doi.org/10.1103/PhysRevC.83.021903}{\doi{10.1103/PhysRevC.83.021903}},
\href{http://www.arXiv.org/abs/1011.3750}{\texttt{ arXiv:1011.3750}}.

\bibitem{Teaney:2010vd}
\hrefCMSnoop {} {D.~Teaney and L.~Yan, ``Triangularity and dipole asymmetry in
  heavy ion collisions'',} \textit{ Phys. Rev. C} \textbf{ 83} (2011) 064904,
  \href{http://dx.doi.org/10.1103/PhysRevC.83.064904}{\doi{10.1103/PhysRevC.83.064904}},
\href{http://www.arXiv.org/abs/1010.1876}{\texttt{ arXiv:1010.1876}}.

\bibitem{ALICE:2011ab}
\hrefCMSnoop {} {{ ALICE} Collaboration, ``{Higher harmonic anisotropic flow
  measurements of charged particles in Pb-Pb collisions at \rootsNN\ = 2.76
  \TeV}'',} \textit{ Phys. Rev. Lett.} \textbf{ 107} (2011) 032301,
\href{http://www.arXiv.org/abs/1105.3865}{\texttt{ arXiv:1105.3865}}.

\bibitem{Khachatryan:2010gv}
\hrefCMSnoop {} {{ CMS} Collaboration, ``Observation of long-range near-side
  angular correlations in proton-proton collisions at the {LHC}'',} \textit{
  JHEP} \textbf{ 09} (2010) 091,
  \href{http://dx.doi.org/10.1007/JHEP09(2010)091}{\doi{10.1007/JHEP09(2010)091}},
\href{http://www.arXiv.org/abs/1009.4122}{\texttt{ arXiv:1009.4122}}.

\bibitem{JINST}
\hrefCMSnoop {} {{ CMS} Collaboration, ``{The CMS experiment at the CERN
  LHC}'',} \textit{ JINST} \textbf{ 0803} (2008) S08004,
  \href{http://dx.doi.org/10.1088/1748-0221/3/08/S08004}{\doi{10.1088/1748-0221/3/08/S08004}}.

\bibitem{EWK-10-004}
\href {http://cdsweb.cern.ch/record/1279145} {{ {CMS}} Collaboration,
  ``Measurement of {CMS} Luminosity'',} CMS Physics Analysis Summary
  CMS-PAS-EWK-10-004, (2010).

\bibitem{EWK-11-001}
\href {http://cdsweb.cern.ch/record/1335668} {{ CMS} Collaboration, ``Absolute
  luminosity normalization'',} CMS Detector Performance Summary
  CMS-DP-2011-002, (2011).

\bibitem{glauber}
\hrefCMSnoop {} {M.~L. Miller {et~al.}, ``{Glauber modeling in high-energy
  nuclear collisions}'',} \textit{ Ann. Rev. Nucl. Part. Sci.} \textbf{ 57}
  (2007) 205,
  \href{http://dx.doi.org/10.1146/annurev.nucl.57.090506.123020}{\doi{10.1146/annurev.nucl.57.090506.123020}}.

\bibitem{Alver:Glauber}
\hrefCMSnoop {} {B.~Alver {et~al.}, ``{The PHOBOS Glauber Monte Carlo}'',}
  (2008). \href{http://www.arXiv.org/abs/0805.4411}{\texttt{ arXiv:0805.4411}}.

\bibitem{Chatrchyan:2011pb}
\hrefCMSnoop {} {{ CMS} Collaboration, ``{Dependence on pseudorapidity and
  centrality of charged hadron production in PbPb collisions at a
  nucleon-nucleon centre-of-mass energy of 2.76 TeV}'',} \textit{ JHEP}
  \textbf{ 08} (2011) 141,
  \href{http://dx.doi.org/10.1007/JHEP08(2011)141}{\doi{10.1007/JHEP08(2011)141}},
\href{http://www.arXiv.org/abs/1107.4800}{\texttt{ arXiv:1107.4800}}.

\bibitem{D'Enterria:2007xr}
\hrefCMSnoop {} {{ CMS} Collaboration, ``{CMS physics technical design report:
  Addendum on high density QCD with heavy ions}'',} \textit{ J. Phys. G}
  \textbf{ 34} (2007) 2307,
  \href{http://dx.doi.org/10.1088/0954-3899/34/11/008}{\doi{10.1088/0954-3899/34/11/008}}.

\bibitem{TRK-10-001}
\href {http://cdsweb.cern.ch/record/1258204} {{ {CMS}} Collaboration,
  ``Tracking and Vertexing Results from First Collisions'',} CMS Physics
  Analysis Summary CMS-PAS-TRK-10-001, (2010).

\bibitem{Lokhtin:2005px}
\hrefCMSnoop {} {I.~P. Lokhtin and A.~M. Snigirev, ``{A model of jet quenching
  in ultrarelativistic heavy ion collisions and high-\pt hadron spectra at
  RHIC}'',} \textit{ Eur. Phys. J. C} \textbf{ 45} (2006) 211,
  \href{http://dx.doi.org/10.1140/epjc/s2005-02426-3}{\doi{10.1140/epjc/s2005-02426-3}},
\href{http://www.arXiv.org/abs/hep-ph/0506189}{\texttt{ arXiv:hep-ph/0506189}}.

\bibitem{star:2007pu}
\hrefCMSnoop {} {J.~Putschke, ``{Intra-jet correlations of high-$p_t$ hadrons
  from STAR}'',} \textit{ J. Phys. G} \textbf{ 34} (2007) S679,
  \href{http://dx.doi.org/10.1088/0954-3899/34/8/S72}{\doi{10.1088/0954-3899/34/8/S72}},
\href{http://www.arXiv.org/abs/nucl-ex/0701074}{\texttt{
  arXiv:nucl-ex/0701074}}.

\bibitem{Agakishiev:2011pe}
\hrefCMSnoop {} {{ STAR} Collaboration, ``{Anomalous centrality evolution of
  two-particle angular correlations from Au-Au collisions at \rootsNN\ = 62 and
  200\GeV}'',}
\href{http://www.arXiv.org/abs/1109.4380}{\texttt{ arXiv:1109.4380}}.

\bibitem{Voloshin:1994mz}
\hrefCMSnoop {} {S.~Voloshin and Y.~Zhang, ``{Flow study in relativistic
  nuclear collisions by Fourier expansion of azimuthal particle
  distributions}'',} \textit{ Z. Phys. C} \textbf{ 70} (1996) 665,
  \href{http://dx.doi.org/10.1007/s002880050141}{\doi{10.1007/s002880050141}},
\href{http://www.arXiv.org/abs/hep-ph/9407282}{\texttt{ arXiv:hep-ph/9407282}}.

\bibitem{Voloshin:2008dg}
\hrefCMSnoop {} {S.~A. Voloshin, A.~M. Poskanzer, and R.~Snellings,
  ``{Collective phenomena in non-central nuclear collisions}'',}
  \href{http://www.arXiv.org/abs/0809.2949}{\texttt{ arXiv:0809.2949}}.

\bibitem{Adare:2010sp}
\hrefCMSnoop {} {{ PHENIX} Collaboration, ``{Azimuthal anisotropy of neutral
  pion production in Au+Au collisions at \rootsNN\ = 200\GeV: Path-length
  dependence of jet quenching and the role of initial geometry}'',} \textit{
  Phys. Rev. Lett.} \textbf{ 105} (2010) 142301,
  \href{http://dx.doi.org/10.1103/PhysRevLett.105.142301}{\doi{10.1103/PhysRevLett.105.142301}},
\href{http://www.arXiv.org/abs/1006.3740}{\texttt{ arXiv:1006.3740}}.

\bibitem{Bass:2008rv}
\hrefCMSnoop {} {S.~A. Bass {et~al.}, ``Systematic comparison of jet
  energy-loss schemes in a realistic hydrodynamic medium'',} \textit{ Phys.
  Rev. C} \textbf{ 79} (2009) 024901,
  \href{http://dx.doi.org/10.1103/PhysRevC.79.024901}{\doi{10.1103/PhysRevC.79.024901}},
\href{http://www.arXiv.org/abs/0808.0908}{\texttt{ arXiv:0808.0908}}.

\bibitem{Peigne:2008wu}
\hrefCMSnoop {} {S.~Peign{\'e} and A.~V. Smilga, ``{Energy losses in a hot
  plasma revisited: QCD versus QED}'',} \textit{ Phys. Usp.} \textbf{ 52}
  (2009) 659,
  \href{http://dx.doi.org/10.3367/UFNe.0179.200907a.0697}{\doi{10.3367/UFNe.0179.200907a.0697}},
  \href{http://www.arXiv.org/abs/0810.5702}{\texttt{ arXiv:0810.5702}}.
Usp. Fiz. Nauk {\bf 179} (2009) 697.

\bibitem{Gubser:2008as}
\hrefCMSnoop {} {S.~S. Gubser {et~al.}, ``{Gluon energy loss in the
  gauge-string duality}'',} \textit{ JHEP} \textbf{ 10} (2008) 052,
  \href{http://dx.doi.org/10.1088/1126-6708/2008/10/052}{\doi{10.1088/1126-6708/2008/10/052}},
\href{http://www.arXiv.org/abs/0803.1470}{\texttt{ arXiv:0803.1470}}.

\bibitem{Wicks:2005gt}
\hrefCMSnoop {} {S.~Wicks\ {et~al.}, ``Elastic, inelastic, and path length
  fluctuations in jet tomography'',} \textit{ Nucl. Phys. A} \textbf{ 784}
  (2007) 426,
  \href{http://dx.doi.org/10.1016/j.nuclphysa.2006.12.048}{\doi{10.1016/j.nuclphysa.2006.12.048}},
\href{http://www.arXiv.org/abs/nucl-th/0512076}{\texttt{
  arXiv:nucl-th/0512076}}.

\bibitem{Marquet:2009eq}
\hrefCMSnoop {} {C.~Marquet and T.~Renk, ``{Jet quenching in the
  strongly-interacting quark-gluon plasma}'',} \textit{ Phys. Lett. B} \textbf{
  685} (2010) 270,
  \href{http://dx.doi.org/10.1016/j.physletb.2010.01.076}{\doi{10.1016/j.physletb.2010.01.076}},
\href{http://www.arXiv.org/abs/0908.0880}{\texttt{ arXiv:0908.0880}}.

\bibitem{Borghini:2000sa}
\hrefCMSnoop {} {N.~Borghini {et~al.}, ``{A New method for measuring azimuthal
  distributions in nucleus-nucleus collisions}'',} \textit{ Phys. Rev. C}
  \textbf{ 63} (2001) 054906,
  \href{http://dx.doi.org/10.1103/PhysRevC.63.054906}{\doi{10.1103/PhysRevC.63.054906}},
\href{http://www.arXiv.org/abs/nucl-th/0007063}{\texttt{
  arXiv:nucl-th/0007063}}.

\bibitem{Borghini:2001vi}
\hrefCMSnoop {} {N.~Borghini {et~al.}, ``{Flow analysis from multiparticle
  azimuthal correlations}'',} \textit{ Phys. Rev. C} \textbf{ 64} (2001)
  054901,
  \href{http://dx.doi.org/10.1103/PhysRevC.64.054901}{\doi{10.1103/PhysRevC.64.054901}},
\href{http://www.arXiv.org/abs/nucl-th/0105040}{\texttt{
  arXiv:nucl-th/0105040}}.

\bibitem{Luzum:2010fb}
\hrefCMSnoop {} {M.~Luzum and J.-Y. Ollitrault, ``{Directed flow at midrapidity
  in heavy-ion collisions}'',} \textit{ Phys. Rev. Lett.} \textbf{ 106} (2011)
  102301,
  \href{http://dx.doi.org/10.1103/PhysRevLett.106.102301}{\doi{10.1103/PhysRevLett.106.102301}},
\href{http://www.arXiv.org/abs/1011.6361}{\texttt{ arXiv:1011.6361}}.

\bibitem{Alver:2008zza}
\hrefCMSnoop {} {B.~Alver {et~al.}, ``Importance of correlations and
  fluctuations on the initial Source Eccentricity in High-Energy
  Nucleus-Nucleus collisions'',} \textit{ Phys. Rev. C} \textbf{ 77} (2008)
  014906,
  \href{http://dx.doi.org/10.1103/PhysRevC.77.014906}{\doi{10.1103/PhysRevC.77.014906}},
\href{http://www.arXiv.org/abs/0711.3724}{\texttt{ arXiv:0711.3724}}.

\end{thebibliography}\endgroup
\cleardoublepage \appendix\section{The CMS Collaboration \label{app:collab}}\begin{sloppypar}\hyphenpenalty=5000\widowpenalty=500\clubpenalty=5000\textbf{Yerevan Physics Institute,  Yerevan,  Armenia}\\*[0pt]
S.~Chatrchyan, V.~Khachatryan, A.M.~Sirunyan, A.~Tumasyan
\vskip\cmsinstskip
\textbf{Institut f\"{u}r Hochenergiephysik der OeAW,  Wien,  Austria}\\*[0pt]
W.~Adam, T.~Bergauer, M.~Dragicevic, J.~Er\"{o}, C.~Fabjan, M.~Friedl, R.~Fr\"{u}hwirth, V.M.~Ghete, J.~Hammer\cmsAuthorMark{1}, M.~Hoch, N.~H\"{o}rmann, J.~Hrubec, M.~Jeitler, W.~Kiesenhofer, A.~Knapitsch, M.~Krammer, D.~Liko, I.~Mikulec, M.~Pernicka$^{\textrm{\dag}}$, B.~Rahbaran, C.~Rohringer, H.~Rohringer, R.~Sch\"{o}fbeck, J.~Strauss, A.~Taurok, F.~Teischinger, P.~Wagner, W.~Waltenberger, G.~Walzel, E.~Widl, C.-E.~Wulz
% ...
\vskip\cmsinstskip
\dag:~Deceased\\
1:~~Also at CERN, European Organization for Nuclear Research, Geneva, Switzerland\\
% ...

\end{sloppypar}
\end{document}